\documentclass[12pt]{article}
\usepackage{amsmath, amssymb, graphics}
\usepackage{amsbsy}
\usepackage{amsfonts}
\usepackage{amsmath}
\usepackage{graphicx}
\usepackage{array}
\usepackage{booktabs}
\usepackage{float}
\usepackage{makecell}
\usepackage[usenames, dvipsnames]{color}

\newcommand{\mathsym}[1]{{}}

\renewcommand{\title}[1]{\vbox{\center\LARGE{#1}}\vspace{5mm}}
\renewcommand{\author}[1]{\vbox{\center#1}\vspace{5mm}}

\makeatletter
\renewcommand\section{\@startsection {section}{1}{\z@}%
                                   {-3.5ex \@plus -1ex \@minus -.2ex}%
                                   {2.3ex \@plus.2ex}%
                                   {\normalfont\large\bfseries}}
\renewcommand\subsection{\@startsection{subsection}{2}{\z@}%
                                   {-3.25ex\@plus -1ex \@minus -.2ex}%
                                   {1.5ex \@plus .2ex}%
                                   {\normalfont\normalsize\bfseries}}
\makeatother
\renewcommand{\theequation}{1.\arabic{equation}}

\renewcommand{\[}{\begin{eqnarray}}
\renewcommand{\]}{\end{eqnarray}}

\newcommand{\cN}{\mathcal{N}}

\def\moth{\mathsurround=0pt}
\newdimen\zo \zo=0pt

\def\tick{\leaders\hrule height 0.5ex depth 0pt \hskip 0.5pt}
\def\upboxfill{$\moth \setbox\zo\hbox{\tick}%
  \hskip 2pt\hbox to 0pt{$\tick$\hss}\hrulefill \hbox to 2pt{$\tick$\hss}$}

\def\dtick{\leaders\hrule height .34pt depth 0.5ex \hskip 0.5pt}
\def\downboxfill{$\moth \setbox\zo\hbox{\dtick}%
  \hskip 2pt\hbox to 0pt{$\dtick$\hss}\hrulefill%
  \hbox to 2pt{$\dtick$\hss}$}

\newcommand{\bea}{\begin{eqnarray} }

\newcommand{\eea}{\end{eqnarray}}
\newcommand{\ee}{\end{equation}}
\newcommand{\be}{\begin{equation}}

\newcommand{\no}{\nonumber}

\def\cN{{\cal N}}

\def\Im{{\rm Im}}

\def\det{{\rm det}}

\def\no{\nonumber}

\def\Tr{{\rm Tr}}

\def\NeqFour{{{\cal N}=4}}

\def\draftnote#1{{\color{red} #1}}

\DeclareMathOperator{\pslash}{\displaystyle{\not}}

\begin{document}
 
\textwidth 170mm
\textheight 230mm
\topmargin -1cm
\oddsidemargin-0.8cm \evensidemargin -0.8cm
\topskip 9mm
\headsep9pt

\overfullrule=0pt
\parskip=2pt
\parindent=12pt
\headheight=0in \headsep=0in \topmargin=0in \oddsidemargin=0in

\vspace{ -3cm} \thispagestyle{empty} \vspace{-1cm}

\begin{flushright}
AEI-2013-262
\hfill
IGC-13/11-02
\end{flushright}

\begin{center}

{\Large \bf Color/kinematics duality for general abelian orbifolds of $\cN=4$ super Yang-Mills theory}

%{\Large \bf with matter in the adjoint and bi-fundamental representation}

\medskip

\vskip .2in

{\large M. Chiodaroli${}^{a, b}$, Q. Jin${}^{a}$ and R. Roiban${}^{a}$}

\bigskip

${}^a${ Institute for Gravitation and the Cosmos \\
 The Pennsylvania State University \\
University Park PA 16802, USA \\ 
{\texttt{qxj103@psu.edu; radu@phys.psu.edu}}
}
\bigskip

${}^b${ Max-Planck-Institut f\"ur Gravitationsphysik,
             Albert-Einstein-Institut\\
Am M\"uhlenberg 1, 14476 Potsdam, Germany\\
\texttt{mchiodar@aei.mpg.de}}

\end{center}

%\vskip .05in

\begin{abstract}

To explore color/kinematics duality for general representations of the gauge group 
we formulate the duality for general abelian orbifolds of the $SU(N)$, $\cN=4$ super 
Yang-Mills theory in four dimensions, which have fields in the bi-fundamental representation, 
and use it to construct explicitly complete four-vector and four-scalar amplitudes at one loop. 
%%In this paper 
%We formulate color/kinematics duality for general abelian orbifolds of
%the $SU(N)$, $\cN=4$ super Yang-Mills theory in four dimensions and use it to construct 
%explicitly complete four-vector and four-scalar amplitudes at one loop. 
%%For fixed amount of supersymmetry  
%%
%
For fixed number of supercharges, graph-organized $L$-loop $n$-point integrands of all orbifold 
theories are given in terms of a  fixed set of polynomials labeled by $L$ representations of the orbifold group. 
In contrast to the standard duality-satisfying presentation of amplitudes of the ${\cal N}=4$ super Yang-Mills theory, 
each graph may appear several times with different internal states.
%
%For fixed number of supercharges,  graph-organized $L$-loop integrands are expressed in terms of a fixed set of polynomials
%labeled by the orbifold group representations of the virtual fields in each loop; in contrast 
%to the standard duality-satisfying presentation of amplitudes of the ${\cal N}=4$ super Yang-Mills theory, 
%each graph may appear several times with different internal states. %The scattering amplitudes of 
%%different theories distinguished by their matter content are specified by the color factor of each graph.
%%
%Employing the color and R-charge flow, we observe that the amplitudes of an orbifold theory can be used to 
%construct the amplitudes of more general quiver gauge theories which reduce to the former for particular 
%choices of couplings, but  do not necessarily exhibit color/kinematics duality on their own.
%
The color and R-charge flow provide a way to deform the amplitudes of  orbifold theories to those of more general 
quiver gauge theories  which  do not necessarily exhibit color/kinematics duality on their own.

Based on the organization of amplitudes required by the duality between color and kinematics in orbifold theories 
we show how the amplitudes of certain non-factorized matter-coupled supergravity theories can be found through 
a double-copy construction.  

%Furthermore, we argue that the organization of amplitudes required by the duality between color and 
%kinematics in orbifold theories can be used to obtain the amplitudes of certain non-factorized matter-coupled 
%supergravity theories through a double-copy construction. 

We also carry out a comprehensive search for  theories with fields solely in the adjoint representation 
of the gauge group  and amplitudes exhibiting color/kinematics duality for all external states and find an interesting 
relation between supersymmetry and existence of the duality.

\end{abstract}

\baselineskip=16pt
\setcounter{equation}{0}
\setcounter{footnote}{0}

\newpage

\tableofcontents

\renewcommand{\theequation}{1.\arabic{equation}}
\setcounter{equation}{0}

\section{Introduction}

Recent detailed investigations of  properties of gauge theories %, with and without matter 
with fields in the adjoint representation of a semi-simple Lie group have revealed that their scattering amplitudes 
have a surprisingly rich structure, especially in the presence of supersymmetry. 
While most of such structure emerges at the planar level, 
color/kinematics (BCJ) duality \cite{Bern:2008qj} relates the leading and subleading color components of scattering 
amplitudes and may extend to the non-planar level some of the remarkable properties of 
%the 
planar amplitudes such as, in the case of $\cN=4$ super Yang-Mills (sYM) theory, 
dual superconformal symmetry \cite{DCI}, the amplitude/Wilson loop duality 
\cite{AmpWL} and the relation between Wilson loops and scattering amplitudes \cite{AmpCor}.

\iffalse

There is by now substantial evidence for the color/kinematics duality in theories with fields in the adjoint representation, especially 
at tree level~\cite{OtherTreeBCJ,BjerrumBohr:2010zs, virtuousTrees,Bern:2010yg,Monteiro:2011pc},
where explicit representations of the numerators in terms of partial
amplitudes are known for any number of external legs~\cite{ExplicitForms}, and  
a partial Lagrangian understanding of the duality is also available~\cite{Bern:2010yg}.  
Additionally, an alternative trace-inspired presentation of color/kinematics duality,
which emphasizes its group-theoretic structure, has been presented in~\cite{Trace}.

At loop level, color/kinematics-satisfying (BCJ form) four-point and five-point amplitudes have been constructed through four \cite{Bern:2010tq} 
and two loops \cite{loop5ptBCJ}, respectively,  in the  $\NeqFour$ sYM theory. In less-than-maximal supersymmetric theories, 
four-point amplitudes have been constructed at one-loop for the  ${\cal N}=1$ and  ${\cal N}=2$ theories in \cite{Carrasco:2012ca}, 
at one and two loops for the pure gauge theory in \cite{Bern:2013yya}, and at one loop with arbitrary multiplicity for the pure   
and $\NeqFour$ sYM theories in \cite{Bjerrum-Bohr:2013iza}.

\fi

There is by now substantial evidence for the duality between the color and kinematic factors (not including propagators)
for the graph-organized integrands of amplitudes of gauge theories with fields in the adjoint representation, both at tree-level 
and at loop level. Moreover, if the integrand of an amplitude is given in a form that manifestly exhibits the duality, 
%color/kinematics  duality, 
then amplitudes in certain factorized supergravity theories can be obtained in the same graph-organized form 
\cite{BCJLoop} by simply replacing the amplitude's color factors with another set of kinematic  factors. 

\iffalse
Moreover, if the integrand of an amplitude is given in a graph-organized form that manifestly exhibits color/kinematics  duality, 
then amplitudes in certain factorized supergravity theories can be obtained in the same graph-organized form 
\cite{BCJLoop} by simply replacing the amplitude's color factors with another set of kinematic  factors. 
\fi
%%%%%%%%%%%%%%%

While of no less importance, four-dimensional field theories with fields in other representations have been comparatively less studied from 
the perspective of their scattering amplitudes.
%\footnote{Recently, the tree-level amplitudes of certain theories with 
%fields in the bi-adjoint representation of a factorized gauge group were discussed in \cite{Cachazo:2013iea}.}. 
%
Among them, quiver gauge theories -- with  product gauge groups and fields in the adjoint and bi-fundamental representations --
are perhaps the simplest and the ones most closely related to theories with only fields in the adjoint representation\footnote{The 
three-dimensional ABJM can be formulated as a quiver gauge theory, with fields in the bi-fundamental representation. Its scattering 
amplitudes have been extensively studied. The formulation of color/kinematics duality, discussed and explored 
in \cite{Huang:2012wr, Huang:2013kca}, was aided by the three-algebra formulation of this theory, in which 
all propagating fields formally carry a single (adjoint-like) color index.}.
Certain quiver gauge theories exhibit a special point in the space of couplings where they become  (regular or non-regular) orbifolds of $\NeqFour$ 
sYM theory.
In this paper we will study theories which have a  ${\bf Z}_n$ orbifold point: we will define and test color/kinematics duality at the orbifold point 
and then argue that amplitudes for a general choice of couplings can be found by simply dressing the amplitudes at the orbifold point 
following the color and R-charge flow.

Compactification of string theory on orbifolds -- smooth spaces modded out by some discrete group $\Gamma$ --
is a classic construction of four-dimensional matter-coupled gauge and gravity theories \cite{orbifolds}. The spectrum of 
(massless) states consists of {\em untwisted}- and {\em twisted}-sector states. The former are the $\Gamma$-invariant 
states of ten-dimensional flat space string theory.  In a closed string theory the latter are zero-length strings which 
are closed up to the action of the orbifold group and thus localized at the orbifold fixed points.
For oriented open strings it is necessary to specify the action of the orbifold group on the Chan-Patton factors, which is 
most conveniently described in terms of $D$-branes, as it was discussed by Douglas and Moore \cite{Douglas:1996sw}. 
Twisted sector states are then described by strings stretched between a stack of $D$-branes and their images under the orbifold group. 
%\draftnote{is this true?}
%
Consequently, these states are massless provided that  the D-branes are placed at a fixed point of the action of the orbifold group;
the corresponding fields transform in the bi-fundamental representation of the gauge group.
This construction can be realized from a field theory perspective \cite{orbifold_qfts} by starting with maximally-supersymmetric 
gauge theory and projecting onto the $\Gamma$-invariant states while allowing $\Gamma$ to act both on the R-symmetry 
and on the gauge group indices. The result is a vast class of quiver gauge theories whose planar limits have special properties. 
If the action of the orbifold group on the gauge degrees of freedom is in a regular representation -- which from a 
string theory perspective is required for the cancellation of tadpoles -- then the ${\cal N}\ge 1$ quiver gauge theories are 
conformal in the multi-color limit. For non-supersymmetric theories conformal invariance is broken at one-loop level in the multi-color 
limit while it is present in the planar theory~\cite{Dymarsky:2005uh}.
In the orbifold theory the couplings of the various gauge-group factors are equal and proportional to that of the parent theory; renormalizability 
however requires that the theory be deformed off this "natural line" to a general quiver theory with the same matter content. 
  
Perhaps the simplest orbifolds are those with trivial action on the gauge degrees of freedom\footnote{Since the planar 
inheritance discussed in \cite{Bershadsky:1998cb} relies on the regularity of the representation of the orbifold group ({\it i.e.} $\Tr[g]=1$ iff $g$ is 
the trivial element of the group), these theories do not exhibit it.}; the resulting  theories are 
${\cal N}=2$ and ${\cal N}=1$ $SU(N)$ sYM theories without additional matter multiplets and  $SU(N)$ gauge theories with zero, two, four 
or six additional scalars  and specific interactions making them the dimensional reduction of $D=6, 8, 10$ pure gauge theories.
Color/kinematics-satisfying representations of four-gluon amplitudes were constructed at one loop in \cite{Carrasco:2012ca} for the former 
and at one and two-loops in \cite{Bern:2013yya} for the latter theories, and were instrumental in obtaining certain amplitudes 
in ${\cal N}\le 4$ supergravity theories with  additional matter multiplets.

In this paper we shall formulate color/kinematics duality for general abelian orbifolds of the $\cN=4$ sYM theory 
and focus on $\Gamma \simeq {\bf Z}_n$.  
%
\iffalse
%An option is to treat separately the fundamental and anti-fundamental indices of a field in the bi-fundamental 
%representation and for each of them to use the commutation relations of the gauge group as Jacobi 
%identities \cite{Henrik}.
A first option is to seek amplitude presentations with graphs in which all internal lines have definite representations, and to use
the commutation relations of the gauge group with generators in the appropriate representation as the starting point for the definition of 
color/kinematics duality \cite{Henrik} .
%
%\footnote{More generally, the 
%commutation relations of the gauge group with generators in a general representation is good starting point for the definition of 
%color/kinematics duality in the presence of fields in such representations.}
%
\fi
%
An option is to seek presentations of amplitudes in which each internal line corresponds to a field in a definite representation of the 
gauge group; then, the commutation relations of the gauge group with generators in the appropriate representation can be interpreted as 
color Jacobi identities and can be used as the starting point for the definition of color/kinematics duality \cite{Henrik}.
Alternatively, the color Jacobi relations relevant to the orbifold theory are taken to be the (appropriately-defined) image of the Jacobi relations
of the parent theory through the projection  \cite{Bershadsky:1998cb} which  truncates it to the daughter theory.  In this second 
approach all 
calculations are effectively done in the parent theory for all except one 
%unspecified 
arbitrarily-chosen propagator for each loop, which is acted upon 
by an orbifold group element; 
the orbifold theory is obtained by summing over all elements of $\Gamma$. 
Since the parent theory is assumed to only have fields in the adjoint representation, its Jacobi relations are the standard ones; however, 
graphs carrying different {\em inequivalent} choices of orbifold group insertion -- either because of a different element of $\Gamma$ or 
because of a different action of a fixed element on the fields running in loops -- are treated independently. As we shall describe in 
section~\ref{orbifold_section}, the kinematic Jacobi relations mix the corresponding kinematic  factors in a pattern determined by the R-charges 
of internal and external legs. 
%%
\iffalse
%%
Alternatively, 
we can realize the orbifold theory by inserting projection operators  in the Feynman graphs of the parent theory \cite{Bershadsky:1998cb}.
In this approach all calculations are effectively done in the parent theory for all except one unspecified propagator for each loop 
which is acted upon by an orbifold group element; the orbifold theory is obtained by summing over all elements of $\Gamma$. 
%
Since the parent theory has only fields in the adjoint representation, Jacobi relations are the standard ones; however, 
graphs carrying different {\em inequivalent} choices of orbifold group insertion -- either because of a different element of $\Gamma$ or because
of a different action of a fixed element on the fields running in loops -- are assigned independent kinematic numerator factors.
%
Due to the change in the position of the orbifold group element required by the color Jacobi relations on the line carrying it, the 
kinematic Jacobi relations will mix different such numerator factors in a pattern determined by the R-charges of internal and external legs. 
%%%
\fi
%%
While we shall adopt the second approach, in section~7 we shall argue that the two definitions of color/kinematics duality described here 
are equivalent for orbifold theories. Thus, for more general quiver gauge theories that do not have an orbifold point as well as for theories 
with fields in other representations one may use the former strategy.  
 
In the framework above, scattering amplitudes  in the orbifold theory are obtained by independently summing all graphs over all orbifold 
group elements inserted 
in each loop.  As we shall see, an interesting feature of this construction is that, for some $\cN\le 1$ amplitudes, the resulting graphs appear to 
have edges corresponding to fields not present in the orbifold theory; such graphs are absent if one does not require that color/kinematics duality 
is present.  While this may appear problematic, all cuts through the "unphysical" propagator(s) vanish.
It should be possible to understand the appearance of such fields from the perspective of a putative Lagrangian whose Feynman graphs produce 
directly amplitudes in a form that exhibit the duality. As discussed in \cite{Bern:2010yg}, such a Lagrangian has only cubic vertices and the vast 
majority of its fields are auxiliary. 

We shall also attempt to classify all field theories with fields in the adjoint representation which exhibit color/kinematics duality for any choice 
of external states and are power-counting renormalizable (though perhaps not actually renormalizable) when reduced to four dimensions. 
%We will find 
%that four- and five-point matter amplitudes uniquely fix them to be either the pure $\cN$-extended sYM theories in various dimensions, or  YM-scalar 
%theories that can be interpreted as the dimensional reduction of a pure gauge theory in higher dimension. 
We will find that they are either the pure $\cN$-extended sYM theories in various dimensions, or YM-scalar theories that can be interpreted as the dimensional reduction of a pure gauge theory in higher dimension; it may also be possible to extend the latter theories with a particular cubic scalar coupling.
In higher dimensions we shall find 
that the tree-level four-fermion amplitude of a YM theory coupled to a single fermion obeys color/kinematic duality only in dimensions $D=3, 4, 6, 10$, 
{\it i.e.} in the dimensions in which the theory is also supersymmetric. In contrast, tree-level four-point amplitudes with at least two external gluons impose 
essentially no constraints as they depend only on the minimal coupling of matter fields and thus are the same in supersymmetric and 
non-supersymmetric theories.
Our results are consistent with \cite{Nohle:2013bfa} where one-loop four-gluon amplitudes have been shown to have a color/kinematic satisfying 
form for general matter content. Similarly to tree-level amplitudes with at most two external matter fields, these amplitudes are insensitive to the
matter self-coupling and thus do not receive contributions from the interactions which may break color/kinematics duality at tree level.
It would be interesting to find ways to avoid these constraints and use the power of color/kinematics duality in theories which may not 
otherwise exhibit it.

The paper is organized as follows. In the next section we review the construction of field theory orbifolds,  and discuss their deformation 
into more general quiver gauge theories.
% and the projection operator that realizes it on all fields and thus on all propagators of Feynman graphs. 
In section~3, after reviewing the color/kinematics duality in theories with fields in 
the adjoint representation of the gauge group and in particular for the $\NeqFour$ sYM theory, we analyze a general $SU(N)$ gauge theory with 
adjoint matter, antisymmetric couplings and cubic and quartic interactions and constrain it such that the four-and five-point amplitudes obey 
%color/kinematics 
the duality.
In section~4 we formulate the duality for a general abelian orbifold at tree- and loop-level, and spell out the kinematic Jacobi relations 
for one-loop amplitudes.
In section~5 we include examples of four-gluon and four-scalar amplitudes in ${\cal N}=2$, ${\cal N}=1$ and ${\cal N}=0$ orbifold quiver 
gauge theories. 
Based on the construction in earlier sections and on the physical interpretation of the kinematic numerator factors we 
discuss in section~\ref{doublecopy} a double-copy-like construction for certain non-factorizable supergravity theories which 
are orbifolds of $\cN=8$ supergravity.
We summarize our results in section~7, comment on their extension to more general (quiver) gauge theories and gauge theories with fields in 
other representations and prove that, for fields in the fundamental representation, our definition of color/kinematics duality reduces to using the 
gauge group defining commutation relations as color Jacobi identities.
% and, using our results in section 3,  comment on the duality properties of gauge theories with general matter couplings 
%at higher loops.
Two appendices contain a summary of our notations and details omitted in section~3.

%\newpage
%
%
%
%
\renewcommand{\theequation}{2.\arabic{equation}}
\setcounter{equation}{0}

%\section{Field theory orbifolds of the $\cN=4$ sYM theory}

\section{Quiver gauge theories and field theory orbifolds \label{sec:orbifolds}}

A general quiver gauge theory is specified by its gauge group factors, the coupling of each factor, and the matter content including 
the representations (adjoint or bi-fundamental) of matter fields  under the gauge group factors and global symmetry groups. 
Particular quiver gauge theories exhibit an "orbifold point" -- {\it i.e.} a particular choice of couplings for which it can be interpreted as 
a field theory orbifold \cite{orbifold_qfts, Bershadsky:1998cb} of some parent theory. 
Orbifold field theories are obtained by consistently truncating a parent field theory to the fields and interactions that are invariant
under some discrete subgroup $\Gamma$ of the global symmetry group. All couplings of the resulting quiver gauge 
theory are equal and are said to be on the "natural line" in coupling space. 
It is worth mentioning that, while the truncation is consistent, the resulting theory may not be renormalizable; to carry out the 
renormalization program it is in principle necessary to deform the theory off the natural line and to allow for different renormalization constants for 
the couplings of different gauge group factors. Generically, the U(1) factor originally accompanying each gauge group acquires 
non-vanishing beta function  \cite{Fuchs:2000mr, Dymarsky:2005nc} and decouples in the IR.

The action of an element $\gamma \in \Gamma$ on the fields of the parent theory is specified by the pair $(r_\gamma,g_\gamma)$ 
giving, respectively, the representation of $\gamma$ in the flavor symmetry group $F$ and in the (global part of) the gauge group $G$. In the
following we will not write explicitly the index $\gamma$ and, with a slight abuse of notation, interpret the elements of the orbifold group as 
%being given by 
the pairs $(r, g)$. In general these representations need not be faithful. 
Perhaps the simplest nontrivial example corresponds to choosing $g=1$, {\it i.e.} a trivial representation of $\Gamma$ in the gauge group;
in these cases, the truncation eliminates some of the fields of the parent theory while preserving the representations of the remaining ones.
Pure $\cN\le 2$ sYM theories can be interpreted as such orbifolds of $\cN=4$ sYM theory.
More interesting theories, with matter fields in the adjoint and bi-fundamental 
representations, are obtained by choosing both $r$ and $g$ to be nontrivial \cite{orbifold_qfts, Bershadsky:1998cb}. 
While in principle one may orbifold any field theory, a judicious choice for the parent theory  and of orbifold group leads to daughter 
theories inheriting interesting properties \cite{Bershadsky:1998cb}.

\iffalse
%%%%%%%%%%%%%%%%%%%
In general, one may pick  $\Gamma$ to act independently on the gauge $G$ and flavor $F$ symmetry groups, {\it i.e.} 
$\Gamma=\Gamma_r\otimes \Gamma_g$ with $\Gamma_r\subset F$ and $\Gamma_g\subset  G$. 
%
The simplest nontrivial examples are theories corresponding to a trivial $\Gamma_g$: in these cases, the truncation eliminates 
some of the fields of the parent theory while preserving the representations of the remaining ones.
More interesting theories are obtained by choosing $\Gamma_g$ to be a particular representation of $\Gamma_r$ in the gauge 
group; an element of the orbifold group is then given by the pair $(r, g_r)$ with $r\in F$ and $g_r\in G$ a representation of 
$r$ in $G$~\footnote{One may choose a trivial representation of $\Gamma$ in $G$ and thus include a trivial $\Gamma_g$ in 
the same framework; more generally, one chooses $\Gamma_g$ to be an unfaithful representation of $\Gamma$ in $G$.}. 
%.
While in principle one may orbifold any field theory, a judicious choice for the parent theory  and the orbifold group leads to the daughter 
theory inheriting interesting properties. 
%%%%%%%%%%%%%%%%%%%
\fi

Well-studied examples \cite{orbifold_qfts} 
are orbifolds of $SU(|\Gamma| N)$ $\NeqFour$ sYM theory with an orbifold group $\Gamma$ of rank $|\Gamma|$
whose elements are pairs $(r, g)$ with $r\in SU(4)$ and $g$ taken to be a faithful and regular representation of $r$ in 
$SU(|\Gamma|N)$.\footnote{Choosing $g$ to be an unfaithful representation of $\Gamma$ leads to inclusion of orbifolds of 
$\cN\le 2$ sYM theories in this framework. However, the interesting properties discussed in \cite{orbifold_qfts, Bershadsky:1998cb} such 
as planar inheritance no longer hold.}

In the following we will assume that $\Gamma$ is abelian and relax the constraints on its representations.
The physical fields of the daughter theory are invariant under the action of all elements of $\Gamma$, \emph{i.e.}
\be 
\Phi_{a_1 \ldots a_n} = r^{a_1}_{a_1} \dots r^{a_n}_{a_n} \ g \Phi_{a_1 \ldots a_n} g^\dagger \label{condorb} \ ,  
\ee
where $a_1, \ldots a_n$ are $SU(4)$ indices in the fundamental representation. Following our assumption that the 
orbifold group is abelian, we have written its generators as diagonal matrices.
It is convenient to introduce explicitly orbifold projection operators \cite{Bershadsky:1998cb} 
which enforce the condition (\ref{condorb}) and act on a generic field as 
\be 
{\cal P}_\Gamma  \Phi^A_{a_1 \ldots a_n} = {1 \over |\Gamma|} \sum_{(r,g)\in \Gamma} r^{a_1}_{a_1} \dots r^{a_n}_{a_n} 
\ g^{AB} \Phi^B_{a_1 \ldots a_n} \ ;
\label{projection_op}
\ee
%where 
the summation is taken over all elements of $\Gamma$.
% and $|\Gamma|$ is the rank of the orbifold group.
In this expression the indices $A$ and $B$ denote an arbitrary representation; for a field in the adjoint representation
they each take $(|\Gamma|^2N^2-1)$ values.
%%%
%In the above expression, we have chosen to associate an adjoint index $A,B=1,2,\ldots |\Gamma|^2N^2-1 $ to the 
%field $\Phi_{a_1 \ldots a_n}$. 
%This  constitutes a particularly convenient choice as amplitude presentations which have manifest 
%color/kinematics duality involve structure constants rather than color traces. 
%
With the normalization $\Tr[T^A T^B] = \delta^{AB} $ we have,
\be 
g^{AB} = \text{Tr} \big( T^A g T^B g^\dagger \big) = (g^\dagger)^{BA} \ . 
\label{defgab} 
\ee

The cases in which $\Gamma$ acts trivially in the gauge group were discussed in detail in \cite{Carrasco:2012ca}:
$\Gamma\subset SU(3)\subset SU(4)$ leads to pure sYM theories and $\Gamma\subset SU(4)$ breaks supersymmetry 
completely and leads to YM theory with $0, 2, 4$ or $6$ complex scalar fields. 
%%%%
%In the very simplest cases, the elements of $\Gamma$ act trivially on the gauge degrees of freedom. 
%The corresponding orbifold theories contain only fields transforming in the adjoint representation of the gauge group. 
%Their amplitudes have already been studied at one loop in \cite{Carrasco:2012ca}, where explicit duality-satisfying representations 
%were constructed in all cases with less-than-maximal supersymmetry.    
%%%%

In general, if the action of the orbifold group in the (parent) gauge group is nontrivial (thought still potentially not 
faithful\footnote{Such cases may be rephrased as orbifolds of a less-than-maximally (s)YM theories.}) the daughter theory 
is a quiver gauge theory with fields transforming in bi-fundamental representations.
A common technical assumption\footnote{For string theory orbifold constructions regularity is necessary for tadpole cancellation.} 
is that the orbifold is \emph{regular}, that is
\be 
\textrm{Tr}g \neq 0 \quad \text{iif} \quad g = 1  \ . 
\label{regular} 
\ee
It was shown in \cite{Bershadsky:1998cb} that, with such an orbifold group, planar scattering amplitudes of the daughter theory 
are inherited from the parent to all orders in perturbation theory. We will not make this assumption, but rather consider a 
general representation of $\Gamma$ in the gauge group; then the parent gauge group $SU(N)$ is broken to $SU(N_1) \times \ldots \times SU(N_n)$
with $N=N_1+\dots +N_n$. 
We will still observe a relation between regularity of the orbifold and absence of tadpole graphs in amplitudes. 

\begin{figure}[tb]
\centering
\includegraphics[scale=0.7]{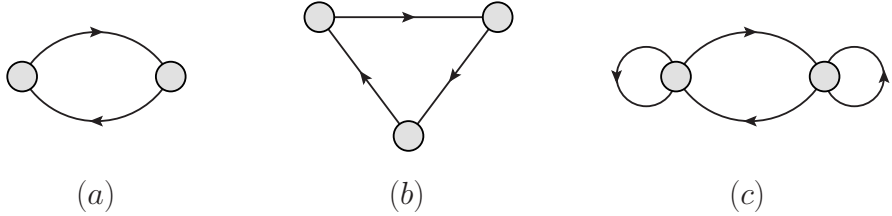}
\caption{Quivers for the $\cN=2, 1, 0$ examples. Each node is a gauge group factor and lines joining them are fields/multiplets 
in bi-fundamental representation. The arrow points from $\bar {\bf N}$ to ${\bf N}$. Lines starting and ending at the same node 
represent adjoint matter fields.}
\label{fig:quivers}
\end{figure}

The simplest non-trivial example, preserving $\cN=2$ supersymmetry, is the regular ${\bf Z}_2$ orbifold generated by 
\be 
r= \text{diag} (1,1,-1,-1) \ , \qquad g = \left( \begin{array}{cc} I_N & 0 \\ 0 & - I_N  \end{array} \right) \ . 
\ee 
This theory has gauge group $SU(N)\times SU(N) \times U(1)$ and contains one $\cN=2$ vector multiplet in the adjoint representation 
of each $SU(N)$ factor, one vector multiplet with the $U(1)$ gauge field and two hypermultiplets transforming in the 
$({\bf N},\bar{\bf N})$ and $(\bar { \bf N},{\bf N})$ representations, respectively. This field content is summarized by the 
quiver in fig.~\ref{fig:quivers}(a).

Similarly, one can obtain an orbifold with $\cN=1$ supersymmetry using the generators 
\be 
r= \text{diag} (1, \omega, \omega, \omega) \ , \qquad g = \left( \begin{array}{ccc} I_N & 0  & 0 \\ 0 &  \omega I_N & 0
\\ 0 & 0 & \omega^2 I_N \end{array} \right) 
\quad \text{with}\quad
\omega^3=1
 \ . 
\ee 
This ${\bf Z}_3$ orbifold theory has gauge group $SU(N)\times SU(N) \times SU(N) \times U(1)^2$. 
The field content amounts to 
five $\cN=1$ vector multiplets and six chiral multiplets. Three of the vector multiplets transform in the adjoint representations of the 
three $SU(N)$ factors, \emph{i.e.} $({\bf N}^2 -1 ,1 ,1 )$, $(1, {\bf N}^2 -1  ,1 )$ and $(1, 1, {\bf N}^2 -1 )$; the remaining two vector multiplets contain the 
$U(1)$ gauge fields. The chiral multiplets transform in $({\bf N}, \bar {\bf N}, 1 )$, $( 1 ,{\bf N}, \bar {\bf N})$ and $( \bar {\bf N}, 1,  {\bf N} )$ 
representations and in the conjugate representations. This field content is summarized by the quiver in fig.~\ref{fig:quivers}(b).

Finally, a simple  ${\bf Z}_2$ orbifold producing an $\cN=0$ theory is generated by
\be 
r= \text{diag} (-1,-1,-1,-1) \ , \qquad g = \left( \begin{array}{cc} I_N & 0 \\ 0 & - I_N  \end{array} \right) \ . 
\ee 
As discussed in \cite{Klebanov:1998yya}, this theory contains the massless modes of a stack of $N$  electric and $N$ magnetic $D3$-branes in 
type 0B string theory.  The gauge group is $SU(N)\times SU(N) \times U(1)$ and 
the field content consists of
one gluon and six adjoint scalars for each $SU(N)$ factor, one U(1) gluon and six additional scalar fields neutral under the $SU(N)\times SU(N)$, 
four fermions transforming in the bi-fundamental representation $({\bf N}, \bar {\bf N})$  and four fermions transforming the anti-bi-fundamental representation  $(\bar {\bf N}, {\bf N})$.  This field content is summarized by the quiver in fig.~\ref{fig:quivers}(c).

\iffalse
In studying the properties of the various orbifold theories, 
one needs to  devote particular care to the treatment of the $U(1)$ factors. As explained in \cite{Fuchs:2000mr} and \cite{Dymarsky:2005nc}, the $U(1)$ 
coupling have generically non-zero beta-functions on the natural line and are driven to smaller and smaller values in the IR  by the RG-flow, so that the $U(1)$ 
fields decouple from the other fields of the theory in the IR.    
\fi

It is not difficult to deform a quiver gauge theory off its orbifold point (if it has one). From a Lagrangian perspective one simply identifies the 
various gauge fields and dresses their interactions with the desired couplings. Similarly, to find the integrands of amplitudes for 
general couplings from those at the orbifold point it suffices to represent them in a cubic graph-based form, which reflects the flow of color 
and R charge. Each vertex of the graph belongs to a single gauge group and thus can be dressed with the desired coupling.

%\newpage

\renewcommand{\theequation}{3.\arabic{equation}}
\setcounter{equation}{0}
\section{Color/kinematics duality for theories with adjoint fields}

\subsection{Review}

The scattering amplitudes of any matter-coupled gauge theory with fields in the adjoint representation and antisymmetric couplings
(an example of which is the $\NeqFour$ sYM theory( can be organized in terms of graphs with only trivalent vertices (cubic graphs); 
assuming that all interactions 
are governed by the gauge coupling $g$, the general expression of the  
dimensionally-regularized $L$-loop  $m$-point scattering amplitude in such a theory is 
%%%
\iffalse
%%
The general expression of the  dimensionally-regularized $L$-loop  $m$-point scattering amplitude in $\NeqFour$ sYM theory (and 
in fact in any field theory with fields in the adjoint representation of a semi-simple gauge group and anti-symmetric structure constant 
couplings) is 
%%
\fi
%%%
\begin{equation}
{\cal A}^{L-\text{loop}}_m\ =\ 
i^L \, g^{m-2 +2L } \,
\sum_{i\in {\cal G}_3}{\int{\prod_{l = 1}^L \frac{d^D p_l}{(2 \pi)^D}
\frac{1}{S_i} \frac {n_i C_i}{\prod_{\alpha_i}{p^2_{\alpha_i}}}}}\, .
\label{LoopGauge} 
\end{equation}
%
%where $g$ is the gauge coupling constant.  
The sum runs over the complete set ${\cal G}_3$ of 
$m$-point $L$-loop cubic graphs, including all permutations of external 
legs, the integration is over the $L$ independent loop momenta $p_l$  
and the denominator is determined by the product of  all propagators of the corresponding 
graph.
The coefficients $C_i$ are the color factors, obtained by assigning to every three-vertex in the 
graph a factor of the antisymmetric structure constant
\be
{\tilde f}^{ABC} = i \sqrt{2} f^{ABC}=\Tr([T^{A},T^{B}]T^{C})\,,
\ee
while respecting the cyclic ordering of edges at the vertex.
The symmetry factors $S_i$ of each graph remove the potential overcount
introduced by the summation over all permutations of external legs included by 
definition in the set ${\cal G}_3$, as well as any symmetries of the graph with fixed external legs.
As in section~\ref{sec:orbifolds}, the gauge group generators $T^A$ are assumed 
to be hermitian and are normalized as $\Tr[T^A T^B] = \delta^{AB}$.
The coefficients $n_i$ are kinematic numerator factors depending on
momenta, polarization vectors and spinors. For supersymmetric
amplitudes in an on-shell superspace they will also contain Grassmann 
parameters.

An amplitude is said to exhibit color/kinematics duality~\cite{Bern:2008qj} if the kinematic 
numerators of a cubic-graph representation of the amplitude satisfy antisymmetry and 
(generalized) Jacobi relations for each propagator, in one-to-one correspondence 
with the properties of color-factors.  That is, for the representation in eq.~\eqref{LoopGauge}, it requires that
\begin{equation}
C_i + C_j + C_k =0 \qquad  \Rightarrow \qquad  n_i + n_j + n_k =0 \, .
\label{BCJDuality}
\end{equation} 
Such representations were conjectured~\cite{BCJLoop} to exist to  all loop orders 
and to all multiplicities in $\NeqFour$ sYM theory; they are related to other representations
by generalized gauge transformations, 
\be
n_i\rightarrow n_i+p_i^2 f(p)~,~~n_j\rightarrow n_j+p_j^2 f(p)~,~~n_k\rightarrow n_k+p_k^2 f(p) \ ,
\label{GenGaugeTransformation}
\ee
which leave the amplitude invariant but reorganize contact terms associated to each graph. Here $f(p)$ can be any function
with the correct dimension and $p_i$, $p_j$ and $p_k$ are the momenta of the internal lines that participate in the Jacobi 
relations \eqref{BCJDuality}.

Color/kinematics duality for pure sYM theories in various dimensions has been discussed extensively, especially 
at tree level~\cite{OtherTreeBCJ,BjerrumBohr:2010zs, virtuousTrees,Bern:2010yg, Trace, Monteiro:2011pc, Cachazo:2013iea, Monteiro:2013rya},
where explicit representations of the numerator factors $n_i$ in terms of color-ordered amplitudes 
are known for any number of external legs~\cite{ExplicitForms, Monteiro:2013rya}. 
Loop-level color/kinematics-satisfying four- and five-point amplitudes have been constructed through 
four-loops \cite{BCJLoop, Bern:2010tq} and two-loops \cite{loop5ptBCJ}, respectively,  in $\NeqFour$ sYM theory. 
In less-than-maximal supersymmetric theories four-point amplitudes have been constructed at one-loop level in  
${\cal N}=1$ and  ${\cal N}=2$ theories \cite{Carrasco:2012ca},  at one and two loops in pure gauge theory 
in \cite{Bern:2013yya}. All-plus one-loop amplitudes with arbitrary multiplicity in pure gauge theory (and, through 
dimension shifting \cite{Bern:1996ja}, one-loop MHV amplitudes $\NeqFour$ sYM theory) have been constructed 
in  \cite{Bjerrum-Bohr:2013iza}.

In the next subsection we will identify all matter-coupled gauge theories with only massless 
fields in the adjoint representation of some semi-simple gauge group and antisymmetric couplings which can obey color/kinematics duality\footnote{We focus on theories that are power-counting renormalizable -- though not 
necessarily renormalizable -- when reduced to four dimensions.}. 
We will find an interesting relation 
with  supersymmetry: whenever pure YM theory coupled to a single fermion is  
supersymmetric in a given dimension $D$,  the corresponding tree-level amplitudes obey color/kinematics duality.
%%%
\iffalse
%
We will then proceed to discuss the duality relations in theories obtained from the ones with fields in the adjoint 
representation through the orbifold projection reviewed in section~2. We will begin at tree-level, where the internal 
states of a graph are uniquely determined by the choice of external legs, and then proceed to loop level. At tree level, 
this construction explains some of the results in section~\ref{sec:adjoint} as inherited from the parent theory.
%
\fi
%%%

%\subsection{Gauge theories coupled with adjoint matter: a general classification}

%\renewcommand{\theequation}{4.\arabic{equation}}
%\setcounter{equation}{0}
\subsection{Color/kinematics duality for gauge theories coupled with adjoint matter: a general classification \label{sec:adjoint}}

%\draftnote{In each of the different cases that are discussed add that 4-point amplitudes with at least 2 gluons are the same as in a 
%susy theory so they obey color/kinematics}

The most general Lagrangian with $n_s$ real adjoint scalars and $n_f$ adjoint fermions which is power-counting renormalizable in four dimensions is
\begin{equation}
\begin{aligned}
{\cal L}=&\Tr\left[-\frac{1}{4}F_{\mu\nu}F^{\mu\nu}-\frac{1}{2}D_{\mu}\phi^ID^{\mu}\phi^I+i\bar{\psi}_A\pslash{D}\psi^A
+\frac{1}{8}\alpha^{IJKL}[\phi^I,\ \phi^J][\phi^K,\ \phi^L]\right.\\
&\left.+\frac{1}{6}\sigma^{IJK}[\phi^I,\phi^J]\phi^K+\frac{i}{\sqrt2}\lambda^I_{AB}\psi^A[\phi^I,\ \psi^B]
+\frac{i}{\sqrt2}\bar{\lambda}^{IAB}\bar{\psi}_A[\phi^I,\ \bar{\psi}_B]\right] \ .\\
\end{aligned}\label{lagrangian}
\end{equation}
Here $\sigma^{IJK}$ and $\alpha^{IJKL}$ are constant coefficients with symmetries dictated by the combination of commutators they multiply.
While the notation might suggest otherwise, we do not assume the existence of any internal global symmetry acting on scalars and fermions.

To test whether this theory can exhibit color/kinematics duality we focus on the four-point amplitudes which probe this unambiguously 
because there is a single Jacobi relation between its numerator factors.
%
%To test whether this theory can exhibit color/kinematics duality we focus on the four-point amplitudes; since generalized gauge transformations
%\eqref{GenGaugeTransformation} act trivially on such amplitudes, they probe unambiguously whether \eqref{lagrangian} can lead to amplitudes
%obeying color/kinematics duality.  
%
Since the four-point amplitudes with at least two external gluons are the same as 
in $\cN=4$ sYM theory (up to the perhaps different number of scalars and fermions)\footnote{Consequently, the generalized unitarity 
method implies that all cuts of one-loop four-gluon amplitudes exhibit color/kinematics duality and thus that the corresponding amplitude 
may exhibit it as well. This is indeed the case, as shown in \cite{Nohle:2013bfa}.}, the first constraints arise from the four-point 
amplitudes with external scalars and fermions.

\subsubsection{The bosonic theory}

We begin by analyzing the bosonic theory in $D$ dimensions. With the Lagrangian \eqref{lagrangian}, the amplitude with four {\it different} 
scalars is 
 \begin{equation}
\begin{aligned}
&\mathcal{A}^\text{tree}_4(1^{\phi^I}2^{\phi^J}3^{\phi^K}4^{\phi^L})
=\alpha^{IJKL}f^{12a}f^{34a}+\alpha^{KIJL}f^{31a}f^{24a}+\alpha^{JKIL}f^{23a}f^{14a}\\
&+\frac{1}{s_{12}}\sigma^{IJM}\sigma^{KLM}f^{12a}f^{34a}+\frac{1}{s_{13}}\sigma^{KIM}\sigma^{JLM}f^{31a}f^{24a}
  +\frac{1}{s_{14}}\sigma^{JKM}\sigma^{ILM}f^{23a}f^{14a} \ .\\
\end{aligned}
\label{notinpairs}
\end{equation}
The origin of each term is clear; requiring that it exhibits a duality between the color and kinematic numerators leads to 
\begin{equation}
\sigma^{IJM}\sigma^{KLM}+\sigma^{KIM}\sigma^{JLM}+\sigma^{JKM}\sigma^{ILM}+s_{12}\alpha^{IJKL}
+s_{13}\alpha^{KIJL}+s_{14}\alpha^{JKIL}=0 \ .
\end{equation}
The terms with different momentum dependence must cancel separately, implying that $\sigma^{IJM}$ obey a Jacobi identity and that 
$\alpha^{IJKL}$ is cyclically invariant in the first three indices. The structure of the Lagrangian \eqref{lagrangian} 
however implies that such a coefficient is 
projected out by the color Jacobi identity. We may therefore set to zero $\alpha^{IJKL}$ with indices not equal in pairs. 

With the notation $\alpha^{IJ}=\alpha^{IJIJ}$, the four-scalar amplitude with pairwise identical scalars is ($I\ne J$)
\begin{equation}
\begin{aligned}
\mathcal{A}^\text{tree}_4(1^{\phi^I}2^{\phi^I}3^{\phi^J}4^{\phi^J})
=&\frac{s_{13}-s_{14}}{2s_{12}}g^2f^{12a}f^{34a}+\alpha^{IJ}\left(f^{13a}f^{24a}+f^{14a}f^{23a}\right)\\
+&\sigma^{IJM}\sigma^{IJM}\left(-\frac{1}{s_{13}}f^{31a}f^{24a}+\frac{1}{s_{14}}f^{23a}f^{14a}\right)\ .
\end{aligned}
\end{equation}
The terms on the second line exhibit color/kinematics duality on their own (as they should, due to the dimensionful nature of $\sigma^{IJK}$)
while a duality between color and kinematics for the terms on the first line requires that
\begin{equation}
\alpha^{IJ}=\frac{1}{2}g^2~,~~(\forall)~I, J \ .
\end{equation}
Thus, the quartic scalar term of \eqref{lagrangian} must be such that it combines with the gauge field into the dimensional reduction 
of a higher-dimensional pure Yang-Mills theory.
It is possible that higher-point tree-level amplitudes in this theory also obey color/kinematics duality.

\iffalse
It is possible to derive further constraints on the theory by examining the five-scalar amplitude
\begin{equation}
\begin{aligned}
\mathcal{A}^\text{tree}_4(1^{\phi^1}2^{\phi^2}3^{\phi^3}4^{\phi^3}5^{\phi^3})
&=g^2\sigma^{123}\left(\frac{(k_1+k_2-k_3)\cdot(k_4-k_5)+s_{45}}{2s_{12}s_{45}}f^{12a}f^{a3b}f^{b45}\right.\\
+&  \left.\frac{(k_2+k_3-k_1)\cdot(k_4-k_5)+s_{45}}{2s_{12}s_{45}}f^{23a}f^{a1b}f^{b45}\right.\\
+&  \left.\frac{(k_3+k_1-k_2)\cdot(k_4-k_5)+s_{45}}{2s_{12}s_{45}}f^{31a}f^{a2b}f^{b45}\right)
+(3\leftrightarrow 4)+(3\leftrightarrow 5)\\
&+(\sigma^{123})^3\frac{1}{s_{13}s_{25}}f^{13a}f^{a4b}f^{b52}+(3,4,5\ \text{permutations}) \ . 
\end{aligned}\label{5phi}
\end{equation}
While the last line obeys color/kinematics duality, the terms proportional to $g^2\sigma^{123}$ do not. Thus, we must either 
require $g=0$ and find the scalar theory of \cite{Du:2011js}  (upon using the fact that $\sigma^{IJK}$ obeys the Jacobi identity 
to set $\sigma^{IJK}=\sigma {\tilde f}^{IJK}$) or set $\sigma^{IJK}=0$ and find the dimensional reduction of YM theory in $D_s=D+n_s$ dimensions.
\fi

\subsubsection{Four-dimensional theories with fermions}

In the absence of additional deformations of the Lagrangian \eqref{lagrangian}, inclusion of fermions coupling to all scalars as in \eqref{lagrangian} 
rules out the bosonic trilinear coupling.  Indeed, the two-scalar-two-fermion amplitude with different scalars,
\begin{equation}
\mathcal{A}^\text{tree}_4(1^{\phi^I}2^{\phi^J}3^{\psi^A}4^{\psi^B})=\frac{[34]}{s_{12}}\sigma^{IJK}\lambda_{AB}^Kf^{12a}f^{34a} 
\quad I\ne J \ ,
\end{equation}
has a single color structure (a second color structure is forbidden by the absence of a $\langle\psi\psi\rangle$ tree-level 
two-point function) and thus cannot exhibit color/kinematics duality. 

If a scalar $\phi^I$ is absent from the Yukawa couplings but interacts with gluons, then there is a one-gluon exchange four-point amplitude
\begin{equation}
\mathcal{A}^\text{tree}_4(1^{\phi^I}2^{\phi^I}3^{\psi^A}4^{\bar{\psi}_A})=\frac{[3|\pslash{k}_1-\pslash{k}_2|4\rangle}{s_{12}}f^{12a}f^{34a} \ ;
\end{equation}
because it has a single color structure, this amplitude also cannot have color/kinematics duality. We therefore conclude that all 
scalars must interact at tree-level  with fermions through Yukawa-type couplings.

To find the constraints on Yukawa couplings we need to examine the four-fermion and other two-scalar-two-fermion amplitudes with different 
scalars:
\bea
\mathcal{A}^\text{tree}_4(1^{\psi^A}2^{\psi^B}3^{\bar{\psi}_C}4^{\bar{\psi}_D})
\!\!\!&=&\!\!\!
-\frac{\langle 34\rangle}{\langle 12\rangle}\lambda^I_{AB}\bar{\lambda}^{ICD}f^{12a}f^{34a}
+\frac{\langle 34\rangle^2}{\langle 13\rangle\langle 24\rangle}g^2\delta_A^C\delta_B^Df^{13a}f^{24a}\cr
&&\qquad -\frac{\langle 34\rangle^2}{\langle 14\rangle\langle 23\rangle}g^2\delta_A^D\delta_B^Cf^{14a}f^{23a} \ ,
\\
\mathcal{A}^\text{tree}_4(1^{\phi^I}2^{\phi^J}3^{\psi^A}4^{\bar{\psi}_B})
\!\!\!&=&\!\!\!
\frac{\langle 14\rangle\langle 24\rangle}{\langle 12\rangle\langle 34\rangle}g^2\delta^{IJ}\delta_A^Bf^{12a}f^{34a}
+\frac{\langle 14\rangle}{\langle 13\rangle}\lambda^I_{AC}\bar{\lambda}^{JBC}f^{13a}f^{24a}\cr
&&\qquad +\frac{\langle 24\rangle}{\langle 23\rangle}\lambda^J_{AC}\bar{\lambda}^{IBC}f^{23a}f^{14a} \ .
\eea
Then, color/kinematics duality requires that
\begin{equation}
\begin{aligned}
\lambda^I_{AB}\bar{\lambda}^{ICD}=&g^2(\delta_A^C\delta_B^D - \delta_B^C\delta_A^D)\equiv g^2\delta^{CD}_{AB}\\
\lambda^I_{AC}\bar{\lambda}^{JBC}+\lambda^J_{AC}\bar{\lambda}^{IBC}=&g^2\delta^{IJ}\delta_A^B \ ;
\end{aligned}
\label{fc}
\end{equation}
in both equations the repeated indices ($I$ and $C$, respectively) are summed over.
To solve these equations we can consider each $\lambda^I_{AB}$ for fixed $A$ and $B$ as a $n_s$ dimensional complex vector; 
there are in all $\frac{(n_f-1)n_f}{2}$ such vectors. The first eq.~\eqref{fc} implies that each of them has norm $g$ and 
they are orthogonal on each other. A solution for $\lambda$ exists only if the number of components of these vectors is larger than the number 
of vectors, {\it i.e.}
\begin{equation}
\label{nsnf1}
n_s\ge \frac{1}{2}(n_f-1)n_f \ .
\end{equation}
A relation between the number of scalars and fermions can be obtained by contracting the bosonic indices
in the second eq.~\eqref{fc} and eliminating the left-hand side using the first eq.~\eqref{fc} with two contracted fermionic 
indices:
\begin{equation}
\label{nsnf2}
n_s=2(n_f-1) \ .
\end{equation}
Equations  \eqref{nsnf1} and \eqref{nsnf2} together imply that 
\begin{equation}
0\le n_f\le 4 \ .
\end{equation}
Curiously, while we may define a four-dimensional 
field theory with an arbitrary number of fermions by dimensionally reducing $D$-dimensional 
YM theory coupled to one fermion, only for $D\le10$ it can exhibit color/kinematics duality. This 
suggests an interesting relation between this duality and supersymmetry.

For $n_f=0,1,2,4$ eqs.~\eqref{fc} can be solved explicitly and have unique solutions while for $n_f=3$ no solution exists 
(see Appendix~\ref{app:solvefc} for details). The resulting Lagrangians are those of ${\cal N}=n_f$ sYM theories or, equivalently,
the  dimensional reduction to four dimensions of minimal sYM theories in $D=4, 6, 10$.

\subsubsection{Single-fermion-coupled Yang-Mills theory in $D$ dimensions}

The results in the previous section suggest that
it is interesting to explore pure gauge theories coupled to a single Majorana fermion in general dimension $D$. 
The relevant Lagrangian is
\begin{equation}
{\cal L}=\Tr\left[-\frac{1}{4}F_{\mu\nu}F^{\mu\nu}+{i\over 2}\bar{\psi}_A\pslash{D}\psi^A\right] \ .
\label{gluon_1fermion}
\end{equation}
It is not difficult to see that the four-gluon and two-gluon-two-fermion amplitudes obey color/kinematics duality.

\begin{figure}[tb]
\centering
\includegraphics[scale=0.75]{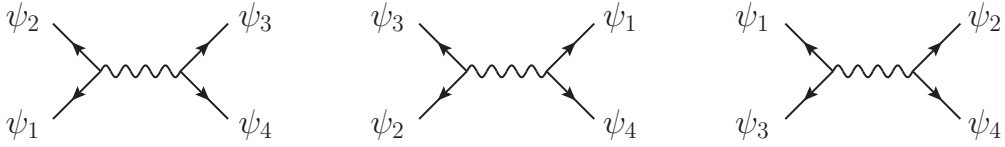}
\caption{The Feynman graphs contributing to the four-fermion amplitude.}
\label{fig:4fermionck}
\end{figure}

The Feynman graphs contributing to the four-fermion amplitude 
%$\mathcal{A}(1^\psi2^{\bar{\psi}}3^\psi4^{\bar{\psi}})$ are 
$\mathcal{A}^\text{tree}_4(1^\psi2^{{\psi}}3^\psi4^{{\psi}})$ are 
shown in fig.~\ref{fig:4fermionck} and the amplitude  is given by
\bea
%\mathcal{A}(1^\psi2^{\bar{\psi}}3^\psi4^{\bar{\psi}})
\mathcal{A}^\text{tree}_4(1^\psi2^{{\psi}}3^\psi4^{{\psi}})
\!\!\!&=&\!\!\!\frac{(\bar{\eta}_1\gamma_{\mu}\eta_2)(\bar{\eta}_3\gamma^{\mu}\eta_4)f^{12a}f^{34a}}{s_{12}}
+\frac{(\bar{\eta}_2\gamma_{\mu}\eta_3)(\bar{\eta}_1\gamma^{\mu}\eta_4)f^{23a}f^{14a}}{s_{14}}\cr
&&\qquad +\frac{(\bar{\eta}_3\gamma_{\mu}\eta_1)(\bar{\eta}_2\gamma^{\mu}\eta_4)f^{31a}f^{24a}}{s_{13}} \ ,
\eea
where $\eta_i$ are spinor external state factors ({\it i.e.} solutions of the free Dirac equation) which obey 
$\bar \eta_i \gamma^\mu \eta_j =  \bar \eta_j \gamma^\mu \eta_i$ due to the Majorana condition $\bar \eta_i = \eta_i^T {\cal C}$. 
%In dimensions in which a Weyl representation
%can be chosen for them the third graph in fig.~\ref{fig:4fermionck} does not contribute and the corresponding third term above vanishes.

The condition that 
%$\mathcal{A}(1^{\psi}2^{\bar{\psi}}3^\psi4^{\bar{\psi}})$ 
$\mathcal{A}^\text{tree}_4(1^{\psi}2^{{\psi}}3^\psi4^{{\psi}})$ obeys color/kinematics duality constrains the external state spinors and 
the Dirac matrices to obey the identity
\begin{equation}
\label{ck10}
(\bar{\eta}_1\gamma_{\mu}\eta_2)(\bar{\eta}_3\gamma^{\mu}\eta_4)+(\bar{\eta}_2\gamma_{\mu}\eta_3)(\bar{\eta}_1\gamma^{\mu}\eta_4)
+(\bar{\eta}_3\gamma_{\mu}\eta_1)(\bar{\eta}_2\gamma^{\mu}\eta_4)=0 \ .
\end{equation}
This analysis can be repeated for pseudo-Majorana spinors which obey the identity $\bar \eta_{ia} \gamma^\mu \eta^a_j =  \bar \eta_{ja} \gamma^\mu \eta^a_i$, 
where the extra indices are contracted with the antisymmetric tensor $\epsilon_{ab}$. We obtain   
\begin{equation}
\label{ck10b}
(\bar{\eta}_{1a}\gamma_{\mu}\eta^a_2)(\bar{\eta}_{3b}\gamma^{\mu}\eta^b_4)+(\bar{\eta}_{2a}\gamma_{\mu}\eta^a_3)(\bar{\eta}_{1b}\gamma^{\mu}\eta^b_4)
+(\bar{\eta}_{3a}\gamma_{\mu}\eta^a_1)(\bar{\eta}_{2b}\gamma^{\mu}\eta^b_4)=0 \ .
\end{equation}

Equations (\ref{ck10}-\ref{ck10b}) are the well-known 
identity that appears in the supersymmetry transformation of the Lagrangian \eqref{gluon_1fermion} and can be 
satisfied only for $D=3, 4, 6, 10$. The appearance of these identities in the color/kinematics relation reinforces the idea that, in the presence of 
fermions, the duality is closely related to existence of supersymmetry.

\iffalse

\subsubsection{Twisted dimensional reduction \draftnote{???? is it worth going into this?}}

\draftnote{do we want to put anything here? The idea would be to somehow forbid the massless mode in the standard dimensional 
reduction and have only a massive mode survive. Then, if color/kinematics survives, construct amplitudes with nonzero mass and take the mass 
to infinity at the end.}

\draftnote{MC: I would just mention this in the conclusion.}

\fi

%\newpage

\renewcommand{\theequation}{4.\arabic{equation}}
\setcounter{equation}{0}
\section{Color/kinematics duality for orbifolds with bi-fundamental fields \label{orbifold_section}}

In this section we define color/kinematics duality for orbifolds with fields in bi-fundamental representations. We will begin by discussing 
tree-level amplitudes and then proceed to loop-level amplitudes. We will then spell out the kinematic Jacobi relations for one-loop amplitudes.
Amplitudes obeying these relations or their higher-loop counterparts are organized in terms of cubic graphs with each edge corresponding 
to a field with definite color and R  charge; thus, each vertex in any given graph is associated to a unique gauge group factor of the orbifold theory.
It is therefore straightforward to obtain the amplitudes of a quiver gauge theory which has the orbifold theory as a special point in its space of
couplings by simply inspecting the color structures  of various graphs and dressing each vertex with the desired coupling constant 
of the corresponding gauge group factor.

\iffalse

In this section we define color/kinematics duality for orbifolds with fields in bi-fundamental representation. We will begin by discussing 
tree-level amplitudes and then proceed to one loop amplitudes. While we will not discuss explicitly higher-loop amplitudes, the relevant 
generalized Jacobi relations can be constructed through similar manipulations.
The results of this section are obtained considering orbifolds in which the couplings corresponding to the different gauge groups are all taken to be equal.
However, it is in principle straightforward to deform a quiver theory away from the orbifold point by inspecting the color structures 
of the various contributions  to the orbifold amplitude and dressing each vertex with the coupling constant of the corresponding gauge group. 
\fi

\subsection{Tree-level amplitudes\label{sectrees}}

It is well-known \cite{Bershadsky:1998cb} that tree-level scattering amplitudes in orbifold field theories can be obtained 
directly from the amplitudes of the parent by simply attaching a projection operator \eqref{projection_op} to each external line.
Indeed, while a projector should formally be included for internal lines as well, their action is trivial as a consequence of the 
external line projection and of the symmetries of the parent theory \cite{Bershadsky:1998cb} which iteratively fix all fields at 
each vertex.

This observation implies that for each internal line of the daughter amplitude there is a color Jacobi identity inherited from the parent by simply 
restricting the color indices to those present in the orbifold theory while not modifying the numerator factors. 
We can do this by introducting a {\it color-space wave functions} $v^A_i$ 
for all external states;  since the orbifold projection correlates the gauge group and R-symmetry (or more generally flavor-symmetry) indices, 
they obey
\be 
v^A_i = R_{i} \  g^{AB} v^B_i , \qquad (\forall) (r,g) \in \Gamma \ , 
%\qquad i=1,2,3,4 \ , 
\label{defv} 
\ee
where $R_{i}$ is the action/representation of a generic orbifold group element $r$ on the $i$-th external particle. 
Without loss of generality, we assume that $\Gamma$ is represented in $SU(4)$ by diagonal matrices; then,
the factor $R_i$ is given by the product of the relevant diagonal entries of $r$: 
\be 
\Phi_i \equiv \Phi_{a_1 \ldots a_n} \ \ \Rightarrow \ \ R_i = r^{a_1}_{a_1} \ldots r^{a_n}_{a_n}  \ .
\ee
For gauge fields, which are uncharged under R symmetry, $R_i=1$ and $v_i^A$ reduce to regular color space wave functions.

Let us illustrate this with a simple four-point tree-level amplitude. A color-dressed four-scalar amplitude in the $\cN=4$ sYM 
theory can be represented as
\be 
{\cal A}_4^{\textrm{tree}} (1^{\phi^{12}},2^{\phi^{23}},3^{\phi^{14}},4^{\phi^{34}}) = g^2 \Big( {c_s n_s \over s } + {c_t n_t \over t } + {c_u n_u \over u }  \Big) \ ,  
\ee
where $g$ is the coupling constant, the upper indices of the arguments of ${\cal A}_4^{\textrm{tree}} $ label the three complex scalars as
the representation ${\bf 6}$ of $SU(4)$ and the color factors are
\be 
%\cN=4 : 
%\quad 
c_s= \tilde f^{A_1A_2 B} \tilde f^{B A_3 A_4} \ , \qquad 
%c_t= \tilde f^{A_2A_3 B} \tilde f^{B A_4 A_1}\ , \qquad 
%c_u= \tilde f^{A_4 A_2 B} \tilde f^{B A_3 A_1} \ .
c_t = \tilde f^{A_1A_4 B} \tilde f^{B A_2 A_3} \ , \qquad 
c_u= \tilde f^{A_1 A_3 B} \tilde f^{B A_4 A_2} \ .
\label{N4c}
\ee
The numerator factors are a solution of the equations
\bea 
{n_s \over s } - {n_t \over t} = - i {t  \over s } \ , 
%\\
\quad
{n_t \over t}- {n_u \over u }  = - i {t \over u} \ ,
%\\
\quad
{n_u \over u}-{n_s \over s }  = - i {t^2 \over s u} \ ,
\eea
and may be obtained through the supersymmetry Ward identities from the corresponding numerator factors of four-gluon amplitudes.

%Orbifolding by a discrete group $\Gamma$ with elements $(r, g)\in SU(4)\times SU(|\Gamma|N)$, the color factors become
Orbifolding by a discrete group $\Gamma$ with elements $(r, g)\in SU(4)\times SU(N)$, 
the color factors become
\bea 
c_s=  v^{A_1}_1 v^{A_2}_2 v^{A_3}_3 v^{A_4}_4 \tilde f^{A_1A_2 B} \tilde f^{B A_3 A_4}  , && \no
%\quad c_t= v^{A_1}_1 v^{A_2}_2 v^{A_3}_3 v^{A_4}_4 \tilde f^{A_2A_3 B} \tilde  f^{B A_4 A_1} , \\
%\quad c_u= v^{A_1}_1 v^{A_2}_2 v^{A_3}_3 v^{A_4}_4 \tilde f^{A_4 A_2 B} \tilde  f^{B A_3 A_1}  , &&
\quad c_t= v^{A_1}_1 v^{A_2}_2 v^{A_3}_3 v^{A_4}_4 \tilde f^{A_1A_4 B} \tilde  f^{B A_2 A_3} , \\
\quad c_u= v^{A_1}_1 v^{A_2}_2 v^{A_3}_3 v^{A_4}_4 \tilde f^{A_1 A_3 B} \tilde  f^{B A_4 A_2}  , &&
\label{orbc}
\eea
where $v_i^{A_i}$ are solutions to the eq.~\eqref{defv} with ($r=\text{diag}(r_1^1, r_2^2, r_3^3, r_4^4)$)
\be
R_{1}=r_{1}^1r_2^2
~,\quad
R_{2}=r_{2}^2 r_{3}^3
~,\quad
R_{3}=r_{1}^{1}r_{4}^{4}
~,\quad
R_{4}=r_3^3r_4^4
\quad \text{with}\quad
r_{1}^1r_2^2r_3^3r_4^4=1 
\label{egR}
\ee
%and a suitable choice of $g$ representing $r$ in $SU(|\Gamma|N)$. 
and a suitable choice of $g$ representing the orbifold group element in $SU(N)$ (and breaking it to $SU(N_1)\times SU(N_2)\times\dots$). 
%\draftnote{MC: it makes sense to call the gauge group of the parent $SU(|\Gamma|N)$ only 
%if we are considering regular orbifolds. Otherwise we should call the gauge group of the parent 
%$SU(N)$ and the group of the orbifold $SU(N_1)\times \dots \times SU(N_k)$
%with $N=N_1 + \dots$.} 
The numerator factors are unchanged.

%It is interesting to 
We note that it is in principle possible that some color factors vanish identically when contracted with the relevant 
color-space wave functions while, as we discussed, the corresponding kinematics numerator factors are unchanged. This does 
not imply a violation of color/kinematics duality since we can assign a non-zero kinematic numerator to the graph with vanishing 
color factor. A similar phenomenon occurs in the color/kinematics-satisfying representation of the four-loop $\NeqFour$ sYM 
superamplitude \cite{Bern:2012uf}, where a vanishing color factor is accompanied by a non-vanishing integrand (which makes a
nontrivial contribution to the corresponding $\cN=8$ supergravity amplitude).

Upon projection to the orbifold invariant states the surviving color space graphs as well as the R charges of fields identify unambiguously 
which gauge group factor governs each cubic vertex; it is therefore straightforward to dress vertices with different couplings for each 
gauge group factor and thus deform the quiver theory off its orbifold point. 
An alternative strategy  \footnote{We thank Lance Dixon for suggesting it.} with the same effect is  
to partition each (tree-level or more generally planar) graph into disconnected sectors which meet on the internal 
and external legs in bi-fundamental representation. All vertices in each such sector belong to a single gauge group and thus 
are dressed with the same coupling.

\subsection{Loop-level amplitudes \label{loop_orb}}

An inspection of the Feynman rules quickly reveals that at loop level  it is possible to remove all but one of the projectors acting 
on internal lines for each independent loop. This should be expected because, unlike tree amplitude, loop amplitudes are not in 
general inherited from the parent theory\footnote{Inheritance is limited to planar amplitudes in theories with a regular 
orbifold action  \cite{Bershadsky:1998cb}.}. 
To  construct Jacobi relations with respect to the projected internal line we begin by making two  observations:
(1) the position of the projection operator is not fixed and can be changed by making use of the $\Gamma$-invariance 
of vertices; (2) while moving the projector from one line to another, terms corresponding 
to different elements of $\Gamma$ are mapped into each other; this is a consequence of {\it e.g.} R-charge conservation at each vertex.

The first observation implies that it is always possible  to make sure that the three graphs related by a color Jacobi relation are 
such that the internal lines participating in the relation are projector-free. The second observation suggests that each color-space
graph with a different insertion of the orbifold group element should be treated as a distinct graph.

In the following we will assign canonically the projector to the internal line carrying the independent loop momentum.
With this labeling the amplitude has the form
\be  
{\cal A}^{(L)} = 
\int \prod_{k=1}^L\frac{d^dl_k}{(2\pi)^d}\,{1\over |\Gamma|}\sum_{(r_k,g_k)\in\Gamma}\;\sum_{R_{l_k} \in {\cal R}}\;
\sum_{i \in {\cal G}_3 } \frac{1}{S_i}{ n_{i; R_{l_1},\dots,R_{l_L}} c_{i; R_{l_1},\dots,R_{l_L}} \over \prod_m p_{m,i}^2} \ , 
\label{OrbifoldLoopGeneral}
\ee
where as in eq.~\eqref{LoopGauge} the summation index $i$ runs over all cubic graphs ${\cal G}_3$ (which includes all possible permutations 
of external legs) and the symmetry factor $S_i$ removes the overcount due to the symmetries of the graph\footnote{Alternatively, one may sum 
only over the inequivalent cubic graphs.}.
$R_{l_1},\dots,R_{l_L}$ are the representations of the orbifold group element $r_1,\dots,r_L$ inside $SU(4)$ corresponding to the fields carrying 
the independent loop momenta $l_1,\dots,l_L$.
The set of all representations that can appear in each loop is denoted by ${\cal R}$.
From a physical perspective, the numerator factor $n_{i; R_{l_1},\dots,R_{l_L}}$ receives contributions from the fields
\footnote{Without imposing color/kinematics duality there are only physical field contributions. However, 
when the duality is imposed, we find that we need to introduce representations which may be related 
to auxiliary fields; this is not surprising from the perspective of a Lagrangian that produces color/kinematics-satisfying 
Feynman rules. All terms in such a Lagrangian are cubic so there are many auxiliary fields. 
See {\it e.g.} \cite{Bern:2010yg} for a few terms in such a putative Lagrangian.} 
with representations 
$R_{l_1},\dots,R_{l_L}$ running in the loop $1,\dots,L$ while the summation over all $R_{l_k}$ is equivalent to the summation over all the fields.
Since $\Gamma$ is assumed to be abelian, $R_{l_k}$ are just phases
% which depend on the particles running around loop 
(see eq.~\eqref{egR} for an example).

The color factors are related -- but not identical -- to the ones of the parent theory: as in the parent theory, to each vertex of the cubic graph 
is assigned a factor of the structure constant of the parent gauge group and their indices are contracted following the edges of the graph:
those corresponding to the edges carrying projectors are contracted with $g_{k}^{AB}$ defined in eq.~\eqref{defgab}, while all the others
with $\delta^{AB}$. Finally, we include an additional overall factor of $\prod_{k=1}^LR_{l_k}$ and a color wave function \eqref{defv} for 
each external leg. It should be noted that, for each choice of orbifold group element, there are as many different color factors as 
elements of ${\cal R}$.
Moreover, the numerator factors depend on the orbifold group element only through 
$R_{l_k} \in {\cal R}$, so that all the numerators corresponding to the same representation of the orbifold group are identical. 

With these preparation we can now describe 
the construction of the kinematic Jacobi relations for an amplitude of the form \eqref{OrbifoldLoopGeneral}:

\begin{enumerate}

\item \label{canonical} Parametrize all graphs by solving momentum conservation and write out the color factors by assigning gauge 
group orbifold elements  $g^{AB}_k$ to the edges carrying the independent loop momenta. The R charge flow is aligned with the momentum 
flow. 

\item \label{start} Choose a graph and an edge of this graph.
% and identify the two other graphs that participate in a color Jacobi relation of the parent theory

\item If this edge does not carry the gauge group orbifold element $g^{AB}$, proceed to the next step. If it does, move it on the 
adjacent edges meeting the chosen one at a vertex using the identity\footnote{This identity can be proven using (\ref{defgab}) to 
show that $g T^A g^\dagger = g^{AB} T^B$ and expressing the structure constants as $\tilde f^{ABC} =\text{Tr}([T^A,T^B]T^C) $.},
\be  
g^{AA'} g^{BB'} g^{CC'} \tilde f^{A'B'C'} = \tilde f^{ABC} 
\quad
\Leftrightarrow
\quad
g^{AA'}  \tilde f^{A'B'C'} = \tilde f^{ABC}(g^\dagger)^{B'B} (g^\dagger)^{C'C}
\label{identityFABC}
\ . 
\ee 
We note that in this equation all $g$ matrices correspond to the same orbifold group element. To help keep track of the $R$ factors 
it is useful to split the  initial $R$ into a product of two factors, each corresponding to the edges carrying the new $g$ factors.

\item \label{use_c_jacobi} 
Use the Jacobi identity of the parent theory for the chosen edge and write the initial color factor  as a sum of color factors associated 
to two other graphs.

\item \label{clean}
Bring the momentum assignment and the two color factors 
to the canonical form  chosen at step \ref{canonical} by repeatedly using the identity \eqref{identityFABC} and the defining property 
of the color wave-functions (\ref{defv}) (or, equivalently, R-change conservation). %When assigning the $R_{l_k}$ factors the orientation of the edge's momentum is important: changing the 
%orientation of the momentum requires inversion of the $R$ factor.\footnote{This is because an out-going particle of some charge 
%$q$ is equivalent to an incoming particle of charge $(-q)$.}

\item \label{kinematic_jacobi} 
The corresponding kinematic Jacobi relation involves the kinematic numerator factors of the original color structure as well 
as of the  two color structures obtained at step \ref{clean}.

\item Go back to step \ref{start}.

\end{enumerate}

Several comments are in order regarding steps \ref{clean} and \ref{kinematic_jacobi}. In the process of rearranging the adjoint orbifold 
group elements at step \ref{clean} several such elements will be multiplied and it may be possible to simplify the product by using the 
defining relations of the orbifold group  ({\it e.g.} for $\Gamma={\bf Z}_n$ and $k<n$ we have $g^{n+k}\simeq g^k$).
One may choose the numerator factor of such a graph in at least two different ways. On the one hand one can use these relations
%the defining properties of the orbifold group 
to simplify all products of orbifold group elements and simply read off the coefficient of the resulting color factor 
coefficient from step \eqref{canonical}. On the other hand, one may interpret $\Gamma$ as part of a (much) larger discrete group 
${\hat \Gamma}$; if the rank of $\hat \Gamma$ is sufficiently large and we seek numerator factors which depend on $\hat \Gamma$ only through 
the representations ${\cal R}$ of $\Gamma$,  the defining relations of the orbifold group need not be used. 
In the latter case, one aims to find a minimal set of representations of the orbifold group ${\cal R}$ for which the numerator factors are 
non-zero and solve the generalized kinematic Jacobi relations. 
We shall choose this second perspective.

%the elements in ${\hat\Gamma}\setminus\Gamma$ are required to have vanishing kinematic numerators; thus, with this
%interpretation, the kinematic factor of a color factor containing $g^{n+k}$ should be taken to be zero. 
%
%%Furthermore, it is in principle possible that there are several  ways to bring to the canonical form the color factors 
%%obtained at step \ref{use_c_jacobi};
%%they can differ only by factors of $g^n$. In this case all kinematic Jacobi relations should hold. 
%%\draftnote{MC: see if the last two paragraphs are OK.}

It is moreover possible that  there are several  ways to bring to the canonical form the color factors obtained at step \ref{use_c_jacobi}; they 
can differ only by elements of $\Gamma$ which are trivial upon use of its defining relations.
% ({\it e.g.} integer powers of $g^n$ for $\Gamma=Z_n$). 
In such cases it is necessary to impose all variants of the kinematic Jacobi relations.

%\draftnote{adjust paragraph above to describe what it is done: that the coefficients with $R$ running over the $Z_n$ range are treated as independent.}

As at tree level, after contraction with the external color wave-functions and summation over the orbifold group elements, it is always possible to use the 
representations of the remaining fields and their R charges to identify the gauge group factor governing each vertex of a given graph. It is then 
straightforward to change the couplings off the natural line and thus find loop amplitudes of the quiver gauge theory at a generic value of its couplings.

Let us now illustrate this construction and write out the kinematic Jacobi relations for the one-loop four-point amplitudes in general 
abelian orbifold theories; we will use them in section~\ref{examples} to construct amplitudes in ${\cal N}=2$, ${\cal N}=1$ and ${\cal N}=0$ orbifold
theories with fields in bi-fundamental representations. To this end we will require the vanishing of 
the numerator factors of graphs containing too high a power of the orbifold group element.

\subsection{A detailed one-loop example  \label{1loop_orb}}

\begin{figure}[t]
\begin{center}
\includegraphics[width=5in]{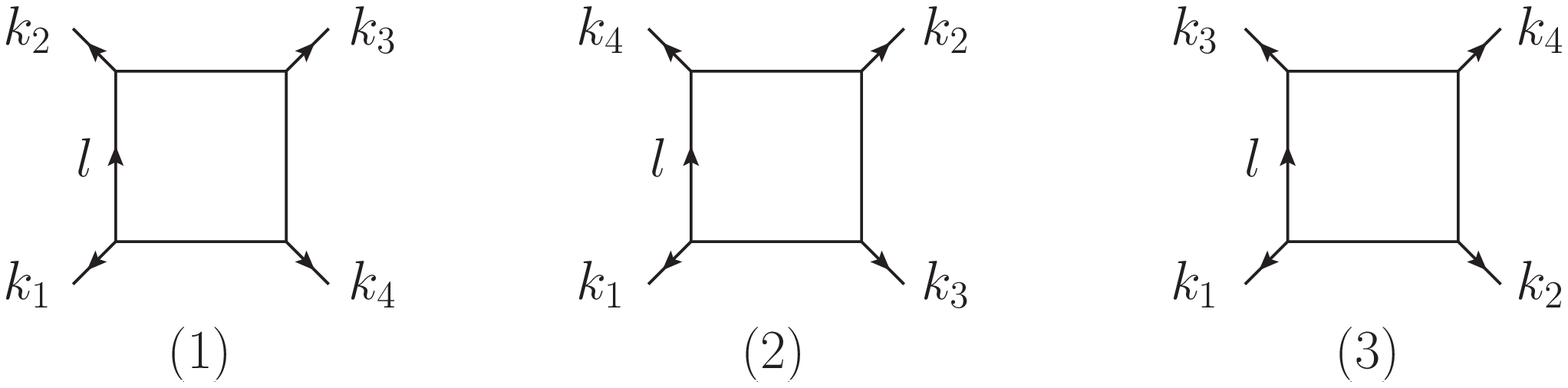}
\caption{Basis of box integrals at one loop. \label{intbox}} 
\bigskip 
\includegraphics[width=5.5in]{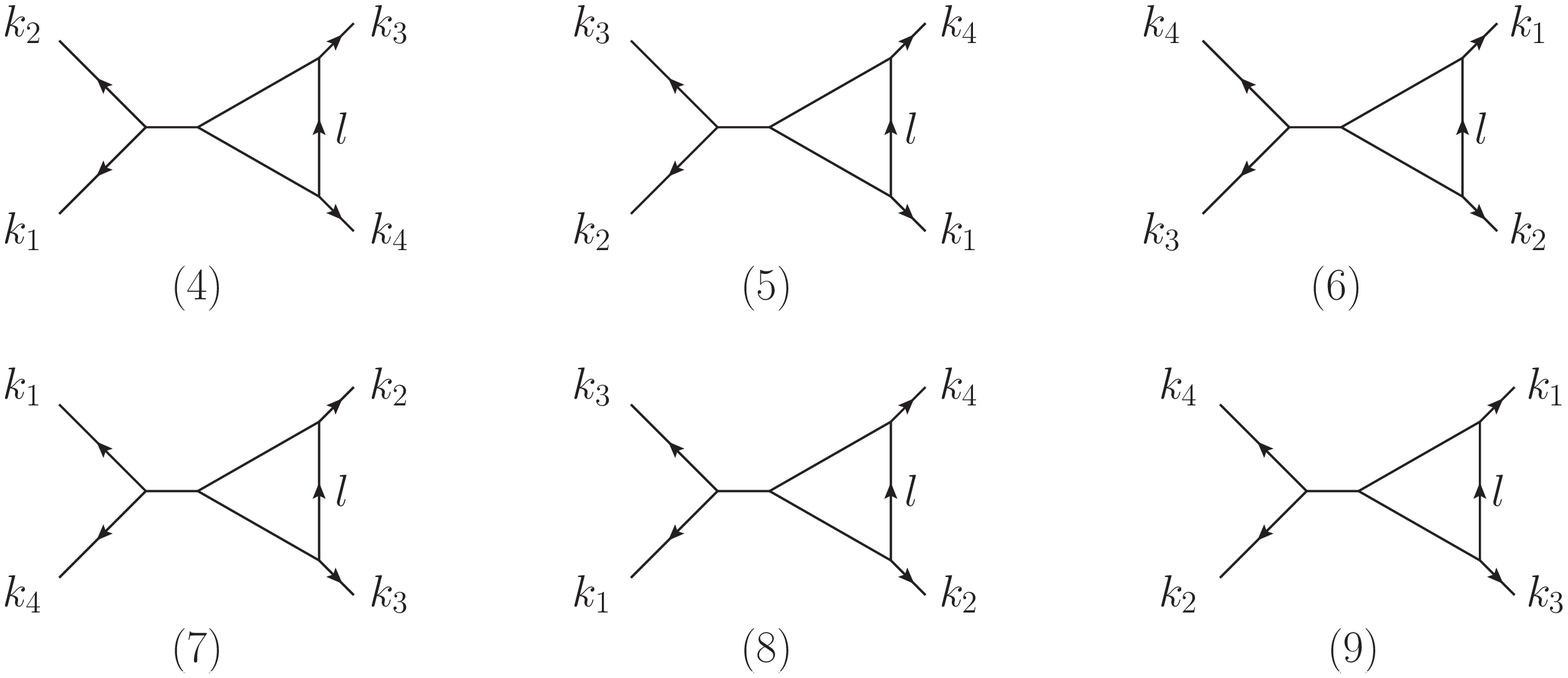}
\caption{Basis of triangle integrals at one loop. \label{inttri}}
\bigskip 
\includegraphics[width=6.1in]{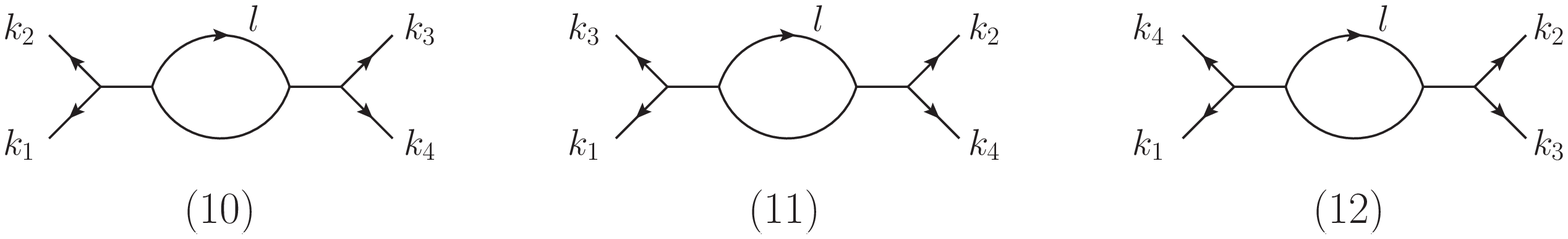}
\caption{Basis of bubble integrals at one loop. \label{intbub}}
\end{center}
\end{figure}
\begin{figure}[t]
\begin{center}
\includegraphics[width=6.3in]{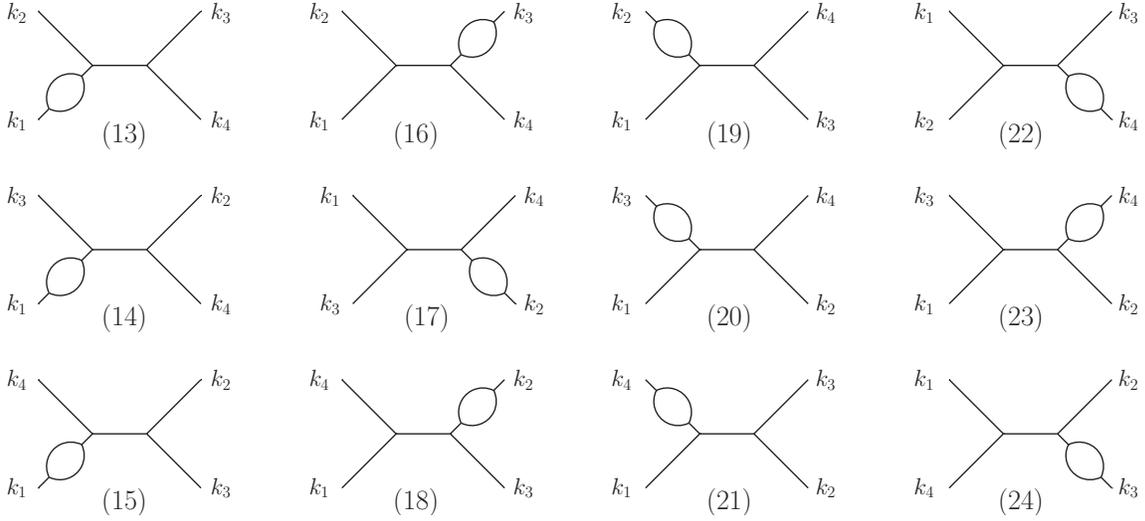}
\caption{Bubble-on-external-line (snail) contributions to one-loop four-point amplitudes. \label{snails}}
\end{center}
\end{figure}

For the one-loop four-point amplitudes
we can choose a basis of cubic graphs with box, triangle and bubble integrals 
shown in figures \ref{intbox}-\ref{intbub};\footnote{There are twelve additional 
bubble-on-external-line ("snail") graphs shown schematically in fig.~\ref{snails} as well as fifteen 
tadpole graphs which we do not draw explicitly. The numerator factors of the former 
will appear in the kinematic Jacobi relations; they are labeled 
from $n_{13} - n_{24}$.} the figures also indicate the internal leg carrying the loop momentum. 
The color factor of each graph is constructed from a structure constant for each vertex  contracted 
with $\delta^{AB}$ or $g^{AB}$. We list here the ones associated to the first, third and fourth graphs: 
\bea 
c_{1;R_{l}} &=& R_{l} \  v^{A_1}_1 v^{A_2}_2 v^{A_3}_3 v^{A_4}_4 \ \tilde f^{A_1A_5A_9}  g^{A_5 A_6}  
\tilde f^{A_2 A_7 A_6}   \tilde f^{A_3A_8A_7} 
\tilde f^{A_4A_9A_8}  \ , 
\label{eg_color1}
\\
c_{3;R_{l}} &=& R_{l} \  v^{A_1}_1 v^{A_2}_2 v^{A_3}_3 v^{A_4}_4 \ \tilde f^{A_1A_5A_9}  g^{A_5 A_6}  
\tilde f^{A_3 A_7 A_6}   \tilde f^{A_4A_8A_7} 
\tilde f^{A_2A_9A_8}  \ , 
\label{eg_color3}
\\
%c_{3;R_{l}} &=& R_{l} \  v^{A_1}_1 v^{A_2}_2 v^{A_3}_3 v^{A_4}_4 \ \tilde f^{A_1A_6A_5}  g^{A_6 A_7}  
%\tilde f^{A_3 A_8 A_7}   \tilde f^{A_4A_9A_8} 
%\tilde f^{A_2A_5A_9}  \ , 
%\\
c_{4;R_{l}} &=& R_{l} \  v^{A_1}_1 v^{A_2}_2 v^{A_3}_3 v^{A_4}_4 \ \tilde f^{A_1A_2A_5}   
\tilde f^{A_5 A_6 A_9}   \tilde f^{A_4A_9A_8} 
g^{A_8A_7 }  \tilde f^{A_3A_7A_6} 
 \ ; 
\label{eg_color4}
\eea
as stated in the previous section, the adjoint  element of the orbifold group $g$ has been inserted on the 
internal line carrying the loop momentum. Also, the direction of the R-charge flow is aligned with the momentum flow. 

To illustrate the construction of the Jacobi identities let us choose the edge carrying the loop momentum in graph 1, {\it i.e.} the line
labeled $A_5$ in eq.~\eqref{eg_color1}. Since this edge also carries $g$ we must move it on the adjacent lines. 
Using eq.~\eqref{identityFABC} we have
%%%%
\iffalse
%%
In order to obtain a set of Jacobi identities suitable for our color factors, 
we have to move $g$ to different internal lines using the invariance of the  structure 
constants under the action of any element of $SU(|\Gamma|N)$\footnote{This identity can be proven using 
(\ref{defgab}) to show that $g T^A g^\dagger = g^{AB} T^B$ and expressing the structure constants as $\tilde f^{ABC} =\text{Tr}([T^A,T^B]T^C) $.},
\be  g^{AA'} g^{BB'} g^{CC'} \tilde f^{A'B'C'} = \tilde f^{ABC} \ . \ee
If we seek the Jacobi identity corresponding to the internal line with color index $A_5$, we can rewrite the color factor as follows,
%%
\fi
%%%%%
\bea 
c_{1;R_{l}} &=& R_{l} \  v^{A_1}_1 v^{A'_2}_2 v^{A_3}_3 v^{A_4}_4 \ (g^{\dagger})^{ A'_2 A_2} (g^{\dagger})^{A_7 A_6} \tilde f^{A_1 A_5 A_9} \tilde f^{A_2  A_6  A_5}  
\tilde f^{A_3 A_8 A_7} \tilde f^{A_4 A_9 A_8} \no \\
&=& {R_{l}\over R_{2}} \  v^{A_1}_1 v^{A_2}_2 v^{A_3}_3 v^{A_4}_4 \  \tilde f^{A_1 A_5 A_9} \tilde f^{A_2 A_6 A_5} g^{A_6 A_7}  \tilde f^{A_3 A_8 A_7}  \tilde f^{A_4A_9A_8} \ .
\label{move_g}
\eea
In the second line we have used the defining property of the color wave-functions (\ref{defv}) as well as 
$(g^{\dagger})^{AB}=g^{BA}$ which is a consequence of the  reality of the adjoint representation. 

The next step is to use the Jacobi relations on the internal line which is now free from the orbifold element $g$:
\bea c_{1;R_{l}} &=& 
{R_{l} \over R_{2}} \  v^{A_1}_1 v^{A_2}_2 v^{A_3}_3 v^{A_4}_4 \ 
( \tilde f^{A_2A_5A_9} \tilde f^{A_1 A_6 A_5} + \tilde f^{A_1A_2A_5} \tilde f^{A_5 A_6 A_9}) g^{A_6 A_7}  
\tilde f^{A_3 A_8 A_7} \tilde f^{A_4 A_9 A_8} \no \\
&=& 
{R_{l} \over R_{2}} \  v^{A_1}_1 v^{A_2}_2 v^{A_3}_3 v^{A_4}_4 \ 
\tilde f^{A_2A_5A_9} \tilde f^{A_1 A_6 A_5}  g^{A_6 A_7}  \tilde f^{A_3 A_8 A_7} \tilde f^{A_4 A_9 A_8}
\cr
&& \qquad
+{R_{l} \over R_{2}R_3} \  v^{A_1}_1 v^{A_2}_2 v^{A_3}_3 v^{A_4}_4 \ 
\tilde f^{A_1A_2A_5} \tilde f^{A_5 A_6 A_9} \tilde f^{A_4 A_9 A_8}(g^\dagger)^{A_8A_7} \tilde f^{A_3 A_7 A_6} 
\label{reorganize_Jacobi}
\eea
where in the triangle graph we have moved the group element past the external line with momentum $k_3$ 
in order to have it in the canonical position, on the line carrying the loop momentum, while keeping the index contraction 
(or equivalently the R-charge flow) as in eq.~\eqref{eg_color4}. In terms of the color factors $c_{i, R}$ eq.~\eqref{reorganize_Jacobi} 
is  
\bea
c_{1;R_{l}} =  c_{3;{ R_{l}/ R_2}} + c_{4;{ R_2 R_3 / R_{l}}}  ,
\label{massaged_color}
\eea
where the inverse $R$ factor compared to \eqref{reorganize_Jacobi} accounts for the presence of $g^\dagger$ in that equation.

%In writing the second color factor we have 
%also inverted the factor $R_l/(R_2R_3)$; this is because the factor in this form corresponds to R-charge flowing {\it out} of the $(k_3, l)$ 
%vertex. Since momentum $l$ is flowing into that vertex, the corresponding field has the opposite charge.

\begin{figure}[t]
\begin{center}
\includegraphics[width=6.3in]{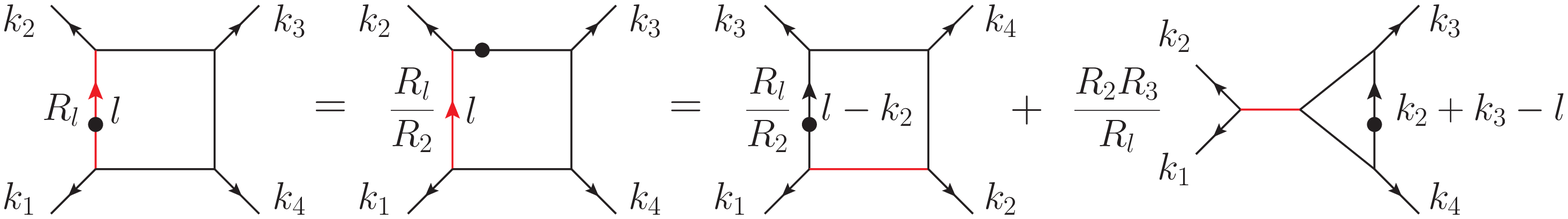}
\caption{Example of Jacobi identity in the orbifold theory. The dot marks the internal line where the adjoint group 
element is inserted. \label{BCJex}}
\end{center}
\end{figure}

The calculation in equations~\eqref{move_g}, \eqref{reorganize_Jacobi} and \eqref{massaged_color} is shown pictorially in fig.~\ref{BCJex}.
From there or from eq.~\eqref{massaged_color} we read off the corresponding kinematic Jacobi relation:
\be 
n_{1; { R_{l}}}\big(l\big) =  n_{3; R_{l}/R_2}\big(l-k_2\big) + n_{4;{R_2 R_3 / R_{l}}}\big(k_2+k_3-l\big) \ . 
\ee 

Repeating the same steps we derive the kinematic Jacobi relations for all the internal edges of all graphs 
in figs.~\ref{intbox}-\ref{intbub}; some of them involve the numerator factors of snail graphs, labeled $n_{13}-n_{24}$ and corresponding 
to the graphs in fig.~\ref{snails}:
\bea 
-n_{1;R_2R_3/R_l}(k_2+k_3-l ) + n_{3;R_3/R_l}(k_3-l ) + n_{4; R_l}(l ) &=& 0 \ , \label{Jacobifirst} \\
n_{4;R_3/(R_4R_l)}(k_3-k_4-l ) + n_{4;R_l}(l ) + n_{10; R_4 R_l}(l + k_4 ) &=& 0 \ , \\
n_{4;R_l}(l ) + n_{4;R_3/R_l}(k_3-l ) - n_{16; R_l/R_3}(l - k_3 ) &=& 0 \ , \\
n_{4;R_l}(l ) + n_{4;1/(R_4R_l)}(-k_4-l ) + n_{22;1/(R_4 R_l)}(-k_4-l ) &=& 0 \ , \\
-n_{1;1/(R_1 R_l)}(-k_1-l ) + n_{2;R_l}(l ) + n_{5; R_l}(l ) &=& 0 \ , \\
n_{5;R_l}(l ) + n_{5;R_4/(R_1R_l)}(k_4-k_1-l ) - n_{12; 1/(R_1 R_l)}( -k_1-l ) &=& 0 \ , \\
n_{5;R_l}(l ) + n_{5;R_4/R_l}(k_4-l ) - n_{21; R_l/R_4}(l - k_4 ) &=& 0 \ ,\\
n_{5;R_l}(l ) + n_{5;1/(R_1R_l)}(-k_1-l ) + n_{15;1/(R_1 R_l)}(-k_1-l ) &=& 0 \ , \\ 
-n_{1;1/ R_l}(-l ) + n_{3;R_l/R_1}(l-k_1 ) + n_{6; R_l}(l ) &=& 0 \ , \\
n_{6;R_1/(R_2R_l)}(k_1-k_2-l ) + n_{6;R_l}(l ) + n_{10; 1/(R_2 R_l)}( -k_2-l ) &=& 0 \ , \\
n_{6;R_l}(l ) + n_{6;R_1/R_l}(k_1-l ) - n_{13; R_l/R_1}(l-k_1 ) &=& 0 \ ,\\
n_{6;R_l}(l ) + n_{6;1/(R_2R_l)}(-l-k_2 ) + n_{19; 1/(R_2R_l)}(-l-k_2 ) &=& 0 \ , \\
-n_{1;R_2/ R_l}(k_2-l ) + n_{2;R_2R_4/R_l}(k_2+k_4-l ) + n_{7; R_l}(l ) &=& 0 \ , \\
n_{7;R_l}(l ) + n_{7;R_2/(R_3R_l)}(k_2-k_3-l ) - n_{12; R_3 R_l}( l+k_3) &=& 0 \ , \\
n_{7;R_l}(l ) + n_{7;R_2/R_l}(k_2-l )  +n_{18; R_l/R_2}(l-k_2 ) &=& 0 \ ,\\
n_{7;R_l}(l ) + n_{7;1/(R_3R_l)}(-k_3-l ) - n_{24; 1/(R_3R_l)}(-k_3-l ) &=& 0 \ , \\
n_{2;R_4/ R_l}(k_4-l ) - n_{3;R_3 R_4/R_l}(k_3+k_4-l ) + n_{8; R_l}(l ) &=& 0 \ , \\
n_{8;R_l}(l ) + n_{8;R_4/(R_2R_l)}(k_4-k_2-l ) - n_{11; R_2 R_l}( l+k_2) &=& 0 \ , \\
n_{8;R_l}(l ) + n_{8;R_4/R_l}(k_4-l )  -n_{23; R_l/R_4}(l-k_4 ) &=& 0 \ ,\\
n_{8;R_l}(l ) + n_{8;1/(R_2R_l)}(-k_2-l ) + n_{17; 1/(R_2R_l)}(-k_2-l ) &=& 0 \ , \\
-n_{2; R_l/R_1}(l-k_1 ) + n_{3;1/R_l}(-l ) + n_{9; R_l}(l ) &=& 0 \ , \\
n_{9;R_l}(l ) + n_{9;R_1/(R_3R_l)}(k_1-k_3-l ) + n_{11;1/(R_3 R_l)}( -k_3-l) &=& 0 \ , \\
n_{9;R_l}(l ) + n_{9;R_1/R_l}(k_1-l )  -n_{14; R_l/R_1}(l-k_1 ) &=& 0 \ ,\\
n_{9;R_l}(l ) + n_{9;1/(R_3R_l)}(-k_3-l ) + n_{20; 1/(R_3R_l)}(-k_3-l ) &=& 0 \ .
\label{Jacobilast} 
\eea

Further identities, relating bubble to tadpole graph numerators, can also be constructed.
To require that the latter numerators vanish identically it suffices to  constrain the numerators of 
the former to obey the identities
\bea 
n_{i; R_l} (l) - n_{i; 1/R_l} (-l) &=& 0 \qquad i\ge 10 \ . 
\label{Jacobitad} 
\eea
In the next section we will use the relations \eqref{Jacobifirst}-\eqref{Jacobilast} and \eqref{Jacobitad} for orbifold groups 
preserving ${\cal N}=2$, ${\cal N}=1$ and ${\cal N}=0$ supersymmetry and for several choices of external states.
It should also be noted that when the external states are taken to be neutral under the orbifold group, the generalized 
Jacobi identities corresponding to the different particles going around the loop decouple and can be solved independently. 
The factors $n_{i;1}$ (and their higher-loop generalizations $n_{i;1\dots 1}$) may be of particular interest as they describe 
the amplitudes of the pure (s)YM theories with the same amount of supersymmetry as preserved by the orbifold group.

We note that the kinematic Jacobi relations -- and consequently the kinematic numerators -- have very limited information 
on the details of the orbifold group; in the color/kinematics-based organization of amplitudes changing the orbifold group (and thus the 
field content of the theory) amounts solely to changing the color factors while keeping the kinematic numerators fixed.

\iffalse
These identities %are the main technical result of the notes, 
can be solved expressing all graphs in terms of three master boxes. 
It is interesting to note that whenever a loop momentum is shifted by one of the external momenta in one of the relations, 
the orbifold element associated to the same external leg also acts on $R_l$.  
\fi

%\newpage

\renewcommand{\theequation}{5.\arabic{equation}}
\setcounter{equation}{0}
\section{Direct computations at one loop \label{examples}}

To construct examples of amplitudes in orbifold theories whose integrands obey the kinematic Jacobi relations constructed 
in the previous section it is perhaps useful to proceed as in the case of the $\NeqFour$ theory and first solve them in terms of 
master graphs. It is not difficult to see that, as in the case of pure ${\cal N}=2$ and ${\cal N}=1$ theories, a possible choice of 
master graphs is given by the box integrals with all possible orbifold group insertions.
This is however not the minimal set. Due to our organization of the calculation -- such that the projection to the orbifold spectrum
is effectively done only upon the  summation over all orbifold group elements -- as well as due to the fact that all information on the 
orbifold group is contained in the color factors, the box integral kinematic numerator factors $n_{1, R_l},n_{2, R_l}$ and $n_{3, R_l}$,  
are closely related to those of the parent ${\cal N}=4$ sYM theory: summing them over all orbifold group elements should yield the 
kinematic numerator factors of a color/kinematic-satisfying representation of ${\cal N}=4$ sYM amplitude.
This constraint determines three box integral numerators in terms of the other ones. In some cases, the number of master graphs can be further reduced by demanding that the amplitudes of a theory with reduced
supersymmetry reproduce as particular cases the known amplitudes of a theory with higher number of supersymmetries. 
 
In each of the explicit calculations that we will discuss we use an ansatz in which the numerator factors are polynomials in the 
Mandelstam variables $s, t$ and $u$ and in the products of external and loop momenta $\tau_{il} = k_i \cdot l$.
When the numerator factors are not expected to be manifestly local ({\it e.g.} due to the presence of polarization vectors in the spinor-helicity basis) inverse powers of the Mandelstam variables are also introduced. 
The degree of the polynomials and the maximum number of loop momenta depend on the residual amount of supersymmetry 
and on the choice of external legs. We then take the following steps to obtain\footnote{The integrands we present in this section 
are correct up to snail integrands and, in the non-supersymmetric examples,  up to rational terms. For massless external 
particles, the snail integrands integrate to zero in dimensional regularization and thus, for $\cN\ge 1$,  the amplitudes we find are complete.}
amplitude presentations with manifest color/kinematics duality.

\begin{enumerate}

\item  Solve (\ref{Jacobifirst}$-$\ref{Jacobilast}); it turns out that imposing absence of tadpoles through eq.~\eqref{Jacobitad}
is not always possible so these equations are not imposed. Construct an ansatz for the master graphs.

\item Fix the free coefficients of the ansatz by imposing that they reproduce the correct $s$-, $t$- and $u$-channel cuts.
To evaluate (generalized) cuts it is useful to use the $\cN=4$ on-shell superamplitudes weighted with the appropriate 
orbifold group elements, as suggested by the inheritance properties of tree-level amplitudes.
%
%For the cut calculation it is useful to use the inheritance properties of tree-level amplitudes, that is, use the on-shell 
%$\NeqFour$ superspace weighted with the appropriate orbifold group elements. 
%
For example, the $s$-channel supercut is given by
\bea 
\!\!\!\!\!
{\cal F}^{1-\text{loop}}_4 \Big|_{s_{12}} \! \! \! \! \! \! &= & \! \! \! \! \! \!  \sum_{(r,g)\in \Gamma}
\! \!  \Big( {\tilde f^{A_1CD} \tilde f^{DBA_2} \over \langle k_2 l_2 \rangle \langle  l_1 k_1 \rangle} -
{\tilde f^{A_1BD} \tilde f^{DCA_2} \over \langle k_2 l_1 \rangle \langle  l_2 k_1 \rangle}\Big) g^{BB'}
\Big( {\tilde f^{A_3B'E} \tilde f^{ECA_4} \over \langle k_4 l_2 \rangle \langle  l_1 k_3 \rangle} -
{\tilde f^{A_3CE} \tilde f^{EB'A_4} \over \langle k_4 l_1 \rangle \langle  l_2 k_3 \rangle}\Big) \no \\
&& \qquad \qquad {\delta^4 \big( \sum_{i=1}^4 \langle l_1 i \rangle \eta^a_i \big) \delta^4 \big( \sum_{i=1}^2 \langle l_2 i \rangle \eta^a_i + 
r^a_a \sum_{i=3}^4 \langle l_2 i \rangle \eta^a_i \big) \over \langle k_1 k_2 \rangle \langle k_3 k_4 \rangle \langle l_1 l_2 \rangle^2   } \ ,
\eea
while the other supercuts can be obtained by relabeling the external legs.
The evaluation of the cut yields a polynomial in the diagonal entries $r_a^a$ of the $SU(4)$ matrices representing the orbifold group. 
Since changing the orbifold group amounts to changing only $r_a^a$, the coefficient of each independent monomial of the cuts of the
ansatz must match the corresponding coefficients in their direct evaluation.

\item Require that in $D$ dimensions the snail integrals do not have any  $1/\mu^2$ pole 
when an infrared regulator $\mu$ is introduced\footnote{This infrared regulator should not be confused with the dimensional regularization 
parameter with a similar notation.}. This condition is necessary to ensure that snail graphs can be included 
in the  presentation of the amplitude (since in these graphs one of the internal lines produces a factor of $1/\mu^2$) and is  
implemented through  the integral reduction
\be  
\!\!
\int \!\! {l^\mu \over l^2 (l+k_i)^2} \rightarrow - {1 \over 2} \int \!\! {k_i^\mu \over l^2 (l+k_i)^2} \ , \ 
\int \!\! {l^\mu l^\nu \over l^2 (l+k_i)^2} \rightarrow {D \over 4(D-1)} \int \!\! {k_i^\mu k_i^\nu \over l^2 (l+k_i)^2} + {\cal O} (\mu^2) \ . 
\label{redsnail} 
\ee
and using some of the free coefficients to set to zero the terms proportional to $1/\mu^2$. 
Alternatively, one may simply fix the contribution of snail graphs by requiring that the amplitude's UV divergence is governed by the beta function(s) 
of the theory and/or that the IR divergences have the expected form, cf. {\it e.g.} \cite{Magnea:1990zb}.

%\draftnote{MC: see if it works now}

\item The remaining free parameters, which drop out upon reduction to master integrals, correspond to either redundancies 
of the ansatz, different representation of the amplitude related by the orbifold version of the generalized gauge 
transformations \cite{BCJLoop}, or to parts of the ansatz that integrate to zero and are not fixed by the standard two-particle cuts. 
Some of them are fixed by requiring manifest Bose/Fermi symmetry of the integrand with respect to permutation of external data. 
Moreover the remaining 
free coefficients are chosen to set to zero the numerators of as many bubble and triangle graphs as possible. 
Last, the residual coefficients can be set to whatever values bring the integrand to one's subjectively chosen simplest form\footnote{Reduction 
to an integral basis can be used to show that the amplitude is independent of these coefficients through ${\cal O}(\epsilon^0)$.}.

\end{enumerate}

\noindent
Let up now follow these steps and construct the four-gluon and four-scalar amplitudes in ${\cal N}=2$,  ${\cal N}=1$ 
and ${\cal N}=0$ supersymmetric orbifold theories.

\subsection{Four-gluon amplitudes}

\subsubsection{Four-gluon amplitudes with $\cN=2$ supersymmetry \label{glueNeq2}}

The theories with the simplest orbifold group action have $\cN=2$ supersymmetry; the case of amplitudes with external gluons is 
particularly easy. Without any loss of generality we take the gluons of momenta $k_1$ and $k_2$ to have negative helicity (and the 
other two of positive helicity) and the diagonal $SU(4)$ matrices $r$ to have the first two entries equal to unity: 
\be
r={\rm diag}(1,1,r_3^3,r_4^4) \ .
\label{Neq2_r}
\ee
Because gluons are uncharged under $SU(4)$, the phase factors $R_i$ representing the orbifold group action on the R-symmetry 
indices of the external lines are trivial,
\be 
R_1=R_2=R_3=R_4 =1 \ . 
\ee
Requiring that $r$ is an $SU(4)$ element, the phase factors $R_l$ that capture the action of the orbifold group on internal lines can be
\be 
R_{l} \in {\cal R} = \big\{ 1,  r^3_3, r^4_4  \big\} \ . 
\ee
Thus, the amplitude contains three copies of each graphs in figures~\ref{intbox}-\ref{intbub}, each dressed with a different color factor, cf. discussion 
in secs.~\ref{loop_orb} and \ref{1loop_orb}.

As mentioned previously, the numerator factors of triangle and bubble graphs are determined in terms of those of box graphs
through the kinematic Jacobi relations (\ref{Jacobifirst}-\ref{Jacobilast}).
Further constraints on the numerator factors $n_{i, r_3^3}$ and $n_{i, r_4^4}$ follow from the $S_2$ subgroup 
of $SU(4)$ that interchanges the R-symmetry indices $3$ and $4$ on all fields. The unit determinant constraint on $r$ is invariant 
under this transformation, which interchanges $r_3^3$ and $r_4^4$. We thus expect that 
\be
n_{i; r^3_3} = n_{i;r^4_4} 
\label{Z2Neq2}
\ee
for all $i=1,\dots,12$. 
%
%%%%%%%%%%
\iffalse
Moreover, as mentioned in the previous section, the fact that the $\cN=4$ sYM theory can be obtained as a 
particular case of the theories (corresponding to the choice $r_3^3=1=r_4^4$) yields the relation
\be 
n_{i;1} + 2 n_{i;r^3_3} = n_i^{\cN=4\;\text{sYM}}, \qquad i=1,2,3 \ ;
\label{reln4} 
\ee
\fi
%%%%%%%%%
Moreover, the fact that $n_{i, R_l}$ receives contributions from fields in the representation $R_l$ of the orbifold group 
implies that, summing over all representations while setting $r=1$ should lead to the numerator factor of the same graph 
in $\cN=4$ sYM theory:
\be 
n_{i;1} + 2 n_{i;r^3_3} = n_i^{\cN=4}, \qquad i=1,2,3 \ ;
\label{reln4} 
\ee
This brings the number of independent master graphs down to three -- the three box graphs with unit orbifold group 
element insertion. 
Last but not least, due to the large amount of supersymmetry preserved by the orbifold group it turns out that it is possible 
to require that two triangle and one bubble diagrams  have vanishing numerator 
factors,
\be 
n_{i;R_l} = 0 \quad \text{for } i=4,6,10, \ (\forall) \ \ R_l \ . 
\ee

We will use the following ansatz for the numerator factors of the three master graphs:
\be 
n_{i;R_l}(l) = i  \hat n_{i;R_l}(l) \langle k_1k_2 \rangle^2 [k_3k_4]^2 , 
\ \ \ \hat n_{i;R_l} = {{\cal P}_{4;1}\big( s, t, \tau_{1l} ,\tau_{2l}, \tau_{3l}, l^2 \big) \over s^2 t u }
+ i {{\cal P}_{2}\big( s, t \big) \epsilon (k_1,k_2,k_4,l) \over s^2 t u }  \ .
\ee
The polynomial ${\cal P}_{4;1}$ is of degree four in all its arguments and of degree up to one in the last four arguments,
$\tau_{il}=k_i \cdot l$ and $l^2$, while ${\cal P}_{2}$ is a polynomial of degree two in its arguments. These polynomials are different for different 
graphs and orbifold phase $R_l$.
As clarified in \cite{Bern:2013yya} for the case of the color/kinematics-satisfying amplitudes in pure $\cN=1$ and $\cN=2$ sYM theories 
found in \cite{Carrasco:2012ca}, the non-locality is presumably due to the choice of helicity basis for the external gluons.  

Matching the $s$-, $t$- and $u$-channel cuts we find that the master graphs' "reduced" numerator factors ${\hat n}_{i, 1}$ are 
\iffalse\bea 
\hat n_{1;1}(l)  \! \! \! \! &= \! \! \! \! & 1 + 2 (\tau_{1l}-\tau_{2l}) { 2 \tau_{3l} - t \over st} + 4{\tau_{2l} (\tau_{2l}+ \tau_{4l})\over tu} \ ,\\ 
\hat n_{2;1}(l) \! \! \! \! &= \! \! \! \! & 1 + 2  {u\over s^2 } \tau_{1l} - 2 {t \over s^2} \tau_{2l}  - 2 {\tau_{3l}\over s } + 4 \tau_{2l} {\tau_{2l}+\tau_{4l}\over t u }
+ 4 \tau_{3l} {\tau_{1l} - \tau_{2 l}\over s t  } + 4 i {\epsilon (k_1,k_2,k_4,l) \over s^2} , \quad \ \ \ \ \\ 
\hat n_{3;1}(l)  \! \! \! \! &= \! \! \! \! & 1 + 2 (\tau_{2l}+\tau_{4l}) { 2 \tau_{2l} - t \over t u} 
+ 4{\tau_{2l} (\tau_{2l}+ \tau_{4l})\over tu} + 4{\tau_{3l} (\tau_{1l} - \tau_{2l})\over st} 
\ .
\eea \fi
\bea 
\hat n_{1;1}(l)   &=   & 1 - {2 \over 3 }  {  \tau_{1l} - \tau_{2l} + l^2 \over s}  \ ,\\ 
\hat n_{2;1}(l)  &=  & 1 + 2 u { \tau_{1l} + \tau_{2l} \over s^2 } + { u + 2\tau_{2l} -2 \tau_{3l} \over 3 s } - {2 l^2 \over 3 s} + 
4 i {\epsilon (k_1,k_2,k_4,l) \over s^2} , \quad \ \ \ \ \\ 
\hat n_{3;1}(l)  &=  & {2 \over 3} - {2 \over 3 }  {  \tau_{1l} + \tau_{2l} + l^2 \over s} 
\ .
\eea

The numerator factors of the box graphs dressed with orbifold group elements follow from eq.~\ref{reln4}
\be
\hat n_{i; r^3_3 }=\hat n_{i; r^4_4 }=\frac{1}{2}(1-\hat n_{i; 1})\qquad i=1,2,3 \ ,
\ee
while the numerator factors for triangle and bubble graphs are determined from the relations \eqref{Jacobifirst}-\eqref{Jacobilast}.
The snail and tadpole graphs have vanishing numerators, as expected. 
The resulting amplitude can be integrated without difficulty, {\it e.g.} by first reducing it to an integral basis and using the known 
expressions for the basis elements.

To obtain the expressions above we have also imposed that the integrand respects Bose symmetry, {\it i.e.} it is 
invariant under the exchange of the external particles of momenta $k_1$ and $k_2$. This further constrains the 
numerator factors to be a solution of
\be 
\hat{n}_{1;R_l}(l) = \hat{n}_{3;R_l}(l-k_1)\big|_{k_1 \leftrightarrow k_2} , \qquad 
\hat{n}_{2;R_l}(l) = \hat{n}_{2;1/R_l}(k_4-l)\big|_{k_1 \leftrightarrow k_2} \label{exchange} \ .
\ee
In general, this exchange symmetry holds for integrated amplitudes. 
We note certain similarities between $\hat n_{1;1}(l), \hat n_{2;1}(l)$ and $\hat n_{3;1}(l)$ and the corresponding 
numerator factors for the four-gluon amplitude in pure $\cN=2$ sYM theory  \cite{Carrasco:2012ca}\footnote{Note that, unlike the one presented in this section, the 
expressions of \cite{Carrasco:2012ca} were obtained setting to zero all three bubble numerators.}. The fact that Bose 
symmetry can also be imposed in that case was pointed out in \cite{Bose_sym} and takes the number of master graphs down to two.

We note that this particular amplitude could have been obtained without any calculation. As we discussed, we choose the master 
integrals to be those corresponding to graphs with the insertion of the unit element of $\Gamma$. In our organization of the amplitude 
\eqref{OrbifoldLoopGeneral} these numerators receive contributions from fields in the trivial representation of $\Gamma$ on $SU(4)$.
These fields are simply those of $\cN=2$ sYM theory, and thus one should have
\be
{\hat n}_{i; 1}={\hat n}_i^{{\cal N}=2} = 1-2{\hat n}_i^{\text{chiral}} 
\qquad
\text{and}
\qquad
\hat n_{i; r^3_3 }=\hat n_{i; r^4_4 } = {\hat n}_i^{\text{chiral}}  \ ,
\ee
where ${\hat n}_i^{\text{chiral}}$ is the contribution of a single chiral multiplet in the loop, denoted by $N_i^\text{chiral}$ in \cite{Carrasco:2012ca}.
%By comparing with \cite{Carrasco:2012ca} it is not difficult to see that this is indeed the case.

\subsubsection{Four-gluons amplitudes with $\cN=1$ supersymmetry \label{glueNeq1}}

Four-gluon amplitudes in $\cN=1$ orbifold theories can be constructed by following the same steps. 
As before, we shall choose the gluons carrying momenta $k_1$ and $k_2$ to have negative helicity and the other two, 
with momenta $k_3$ and $k_4$, to have positive helicity. Without loss of generality we can choose the representation of the 
orbifold $r$ matrix inside the R-symmetry group, $r\in SU(4)$, to be
\be
r={\rm diag}(1, r_2^2, r_3^3, r_4^4) \ ,
\ee
where the nontrivial entries are related by the unit determinant condition. The phases representing the action of
the orbifold group elements on the external lines are trivial,
\be
R_1=R_2=R_3=R_4 =1 \ ;
\ee
the internal lines can be dressed with the phases
\be 
R_{l} \in {\cal R} = \big\{ 1, 
r^2_2, r^3_3, r^4_4, 
r^2_2 r^3_3, r^2_2 r^4_4, r^3_3 r^4_4 \big\} \ . 
\label{Neq1_gluon_phases}
\ee

Despite the reduced amount of supersymmetry, it is still possible to set to zero the snail numerator factors.

The $S_3$ subgroup of the $SU(4)$ R-symmetry of the parent theory which permutes the nontrivial entries of the orbifold matrix 
suggests that we can choose 
\be 
n_{i; r^3_3} = n_{i;r^4_4}  = n_{i; r^2_2} \ , \qquad  n_{i; r^2_2 r^4_4} = n_{i; r^3_3 r^4_4} = n_{i;r^2_2 r^3_3}  \ .
\label{n1-symmetries}
\ee
Moreover, to recover the $\cN=4$ sYM theory for a particular choice or orbifold ($\Gamma={\bf 1}$, $r_2^2=r_3^3=r_4^2=1$, {\it etc}), 
the numerators should obey 
\be 
n_{i;1} + 3 n_{i;r^2_2} + 3 n_{i;r^2_2 r^3_3} = n_i^{\cN=4} \ , \qquad i=1,2,3 \ . 
\label{reln4-n1} \ee
There are therefore six independent numerators. We will choose the three box integrals   with $R_l=1$ and $R_l=r_2^2$ as master graphs
and parametrize their numerators as
\be 
n_{i;R_l}(l) = i  \hat n_{i;R_l}(l) \langle k_1k_2 \rangle^2 [k_3k_4]^2, \quad
\hat n_{i;R_l}(l) = {{\cal P}_{5;2}\big( s, t, \tau_{il}, l^2 \big) \over s^3 t u }
+ i {{\cal P}_{3;1}\big( s, t , \tau_{il}  \big) \epsilon (k_1,k_2,k_4,l) \over s^3 t u } , 
\ee
where ${\cal P}_{5;2}$ and ${\cal P}_{3;1}$ are, respectively, polynomials of degree five and three which also have at most degrees two 
and one in their loop-momentum-dependent arguments. As in the $\cN=2$ theories, they are different for different 
graphs and orbifold phase $R_l$. Matching the generalized cuts and setting the remaining coefficients as 
explained below leads to 
 \bea 
 \label{Neq1n11}
 \hat n_{1;1}(l)  \! \! &= \! \! & 1 +  {\tau_{2l}-\tau_{1l} - l^2 \over s} \ ,\\ 
\hat n_{1;r^2_2}(l) \! \! &= \! \! & - (\tau_{1l}+\tau_{2l}) \Big({t u\over 4 s^3}+ {\tau_{1l}-\tau_{2l} \over s^2} + {l^2 \over s^2} \Big) 
+ {3 \tau_{1l} + 2 \tau_{2l} \over 3 s}
+ {l^2 \over 6 s }\ , \\
\hat n_{2;1}(l)  \! \! &= \! \! & 1 + {u \over 2 s} + 3 u  {\tau_{2l}+\tau_{1l}\over s^2} +  {\tau_{2l}-\tau_{3l}\over s} - {l^2 \over s} + 6 i {\epsilon({k_1,k_2,k_4,l})\over s^2}  \ ,\\ 
\hat n_{2;r^2_2}(l) \! \! &= \! \! &  - {u \over 4 s} +  {t u^2 \over 8 s^3}+ (\tau_{1l}+\tau_{2l}) \Big({u \over 2 s^2 } + { 2 \over 3 s } - { t u  \over  4 s^3} 
- {l^2 \over s^2} + 4 i {\epsilon(k_1,k_2,k_4,l) \over s^3} \Big)
\no \\  
  \! \! & \! \! &  - {\tau_{2l} + \tau_{4l} \over 6 s} +  (\tau_{1l}+\tau_{2l})^2 {u -t \over s^3} + 2{ \tau_{1l}\tau_{4l} \over s^2} +{2u - t \over 6 s^2} l^2 \ , \\
  \hat n_{3;1}(l)  \! \! &= \! \! & {1 \over 2 } - {\tau_{1l} +\tau_{2l} + l^2 \over s} \ ,\\ 
\hat n_{3;r^2_2}(l) \! \! &= \! \! &  \Big( {1 \over 2} + { \tau_{1l}+\tau_{2l} \over s} \Big) 
\Big( {1 \over 2} - {t u \over 4 s^2} - { \tau_{1l}+\tau_{2l} + l^2 \over s} \Big) +  {l^2 \over 6 s } \ . 
 \label{Neq1n3r22}
\eea
As in the $\cN=2$ case,  the numerator factors for triangle graphs can be obtained from eqs.~\eqref{Jacobifirst}-\eqref{Jacobilast}, 
while the numerator factors, $\hat n_{1;r^2_2 r^3_3}=\hat n_{2;r^2_2 r^3_3}=\hat n_{3;r^2_2 r^3_3}$, of the remaining box integrals  
follow from eq.~\eqref{reln4-n1}.
To obtain eqs.~\eqref{Neq1n11}-\eqref{Neq1n3r22} some of the Ansatz' coefficients not determined by the generalized cuts have need 
fixed such that the integrand is manifestly Bose-symmetric under in the external lines $1$ and $2$, as in eq.~(\ref{exchange}).
We have also been able to set to zero one bubble, two triangle and all snail graphs,
\be 
\hat n_{i;R_l} \equiv 0 \ , \qquad (\forall) \ R_l \ , \ \quad i=4,6,10  \quad \text{and} \quad i>12  \ , 
\ee 
while the remaining free coefficients have been set to zero for simplicity. 

It is worth mentioning that, unlike the $\cN=2$ case, the known four-gluon amplitudes of $\cN=2$ and $\cN=1$ pure sYM theories in BCJ form 
do not completely determine the master graph numerators in the orbifold theory. 
Indeed, using the fact that the invariant spectrum is that of $\cN=1$ sYM theory and that for $r_1^1=1$ the invariant spectrum is 
that of $\cN=2$ sYM theory we also have the relations
\be
n_{i; 1}=n_i^{\cN=1}
\quad
\text{and}
\quad
n_{i; 1} + n_{i; r_1^1} + n_{i; r_3^3r_4^4} = n_i^{\cN=2} \ .
\ee
However, $n_{i; r_1^1}$ and $n_{i; r_3^3r_4^4}$ appear in these equations in the same way as in \eqref{reln4-n1} (cf. eq.~\eqref{n1-symmetries})
and thus only their sum is fixed.

\subsubsection{Four-gluon amplitudes with no supersymmetry}

As in the previous cases, the gluons carrying momenta $k_1$ and $k_2$ are taken to have negative helicity and the gluons 
with momenta $k_3$ and $k_4$ are taken to have positive helicity. 
%We now need the  orbifold $r$ matrix inside the R-symmetry group, $r\in SU(4)$, to be completely general
The orbifold  $r$ matrix is now a general diagonal element of $SU(4)$,
\be
r={\rm diag}(r^1_1, r_2^2, r_3^3, r_4^4) \ ,
\ee
with the entries are related by the unit determinant condition. The phases representing the action of
the orbifold group elements on the external lines are trivial as before,
\be
R_1=R_2=R_3=R_4 =1 \ ;
\ee
the internal lines can be dressed with the phases
\be 
R_{l} \in {\cal R} = \big\{ 1, 
r^1_1, r^2_2, r^3_3, r^4_4, 
r^1_1 r^2_2, r^1_1 r^3_3, r^1_1 r^4_4 , r^2_2 r^3_3, r^2_2 r^4_4, r^3_3 r^4_4, r^1_1 r^2_2 r^3_3 , r^1_1 r^2_2 r^4_4, 
r^1_1 r^3_3 r^4_4, r^2_2 r^3_3 r^4_4 \big\} \ . 
\label{Neq0_gluon_phases}
\ee

The $S_4$ subgroup of the $SU(4)$ R-symmetry of the parent theory which permutes the nontrivial entries of the orbifold matrix 
suggests that we can choose 
\bea 
&& n_{i; r^3_3} = n_{i;r^4_4}  = n_{i; r^2_2} = n_{i; r^1_1} \ , \no \\ 
&& n_{i; r^2_2 r^4_4} = n_{i; r^3_3 r^4_4} = n_{i;r^2_2 r^3_3} = n_{i;r^1_1 r^3_3} =  n_{i;r^1_1 r^4_4} = n_{i;r^1_1 r^2_2}   \ , \no \\
&& n_{i; r^1_1 r^2_2 r^4_4} = n_{i; r^1_1 r^3_3 r^4_4}  = n_{i; r^2_2 r^3_3 r^4_4} = n_{i; r^1_1 r^2_2 r^3_3} \ .
\eea
Moreover, to recover the $\cN=4,2,1$ sYM theories for  particular choices of orbifold group, 
the numerators should obey the following relations ($i=1,2,3$)
\bea 
n_{i;1} + 4 n_{i;r^1_1} + 6 n_{i;r^1_1 r^2_2} + 4  n_{i;r^1_1 r^2_2 r^3_3} &=& n_i^{\cN=4} \label{condN4} \ , \\ 
 n_{i;1} + 2 n_{i;r^1_1} + 2 n_{i;r^1_1 r^2_2} + 2  n_{i;r^1_1 r^2_2 r^3_3} &=& n_{i,1}^{\cN=2} \label{condN2} \ , \\
 n_{i;r^1_1} + n_{i;r^1_1r^2_2}  &=& n_{i,r^2_2}^{\cN=1}  \label{condN1} %\  , \qquad i=1,2,3 
 \ .  
\eea 
Taking in account these relations, we can choose the first two boxes with $R_l=1$ and $R_l=r_1^1$ as master graphs
and parametrize their numerators as
\be 
n_{i;R_l}(l) = i  \hat n_{i;R_l}(l) \langle k_1k_2 \rangle^2 [k_3k_4]^2, \quad
\hat n_{i;R_l}(l) = {{\cal P}_{6;3}\big( s, t, \tau_{il}, l^2 \big) \over s^4 t u }
+ i {{\cal P}_{4;2}\big( s, t , \tau_{il}  \big) \epsilon (k_1,k_2,k_4,l) \over s^4 t u } , 
\ee
where ${\cal P}_{6;3}$ and ${\cal P}_{4;2}$ are, respectively, polynomials of degree six and four which also have at most three and two powers of the loop momentum $l$. 

Generalized unitarity fixes the 
%independent 
numerator of the first box in fig.~\ref{intbox} to be,
\bea \hat n_{1;1} & = &  1 + 4 {\tau_{1l} + 3 \tau_{2l} - l^2 \over 3 s} \ ,  
%\\
%\hat n_{1; r^1_1} & = &   - \tau_{1l} \Big( { 1\over 3s } + {t u \over 4 s^3} \Big)  - \tau_{2l} \Big( { 2\over 3s } + {t u \over 4 s^3} \Big)
%+ {\tau_{2l}^2 - \tau_{1l}^2 \over s^2} + l^2 \Big( {1 \over 6s} -{\tau_{1l}  + \tau_{2l} \over s^2} \Big) \ . \quad
 \eea
Additionally, the 
%independent 
numerator of the second box in fig.~\ref{intbox} is
\bea \hat n_{2;1} \! \!  &  \! \!  =  \! \! & \! \!  {u-t \over s} + 8{ u \tau_{1l}^3 - t \tau_{2 l}^3 \over s^4} +  { 4 \tau_{4l} \over 3 s} - 
4 \tau_{2l}^2 \Big( {t \over s^3} +  {2 t \tau_{4l} \over s^3 u}  \Big) + 4 \tau_{1l}^2 \Big( {u\over s^3} - 2{t-2u \over s^4} \tau_{2l} + {2 u \tau_{4l} \over s^3 t} \Big)
\no \quad \\
\! \!  &  \! \!   \! \! & \! \! - 4 \tau_{1l} \Big( {t \over s^2} + {t-u \over s^3} \tau_{2l} + 2 {2t - u \over s^4} \tau_{2l}^2  + {2 \tau_{4l}^2 \over s^2 t}\Big)+
4 \tau_{2l} \Big( {u \over s^2} + {2 \tau_{4l}^2 \over s^2 u} \Big)  \quad \no\\
\! \!  &  \! \!   \! \! & \! \! - 4 l^2 \Big( {1 \over 3 s } + {u \tau_{1l} - t \tau_{2l}\over s^3} \Big) + 8 i \Big( {1 \over s^2} + {\tau_{1l}+ \tau_{2l} \over s^3} +
2{(\tau_{1l}+ \tau_{2l})^2 \over s^4}\Big) \epsilon (k_1,l_2,k_4, l) \ ,
%\no 
%\\
%
% \\
%
%\hat n_{2;r^1_1} \! \!  &  \! \!  =  \! \! & \! \!  {tu^2 \over 8 s^3}  - {11 u \over 12 s} - 4{ u \tau_{1l}^3 - t \tau_{2 l}^3 \over s^4}  - 
% \tau_{2l}^2 \Big( {1 \over s^2} -  {4 t \tau_{4l} \over s^3 u}  \Big) + \tau_{1l}^2 \Big( {1\over s^2} + 4{t-2u \over s^4} \tau_{2l} - {4 u \tau_{4l} \over s^3 t} \Big)-
%\no \ \\
%\! \!  &  \! \!   \! \! & \! \!  \tau_{1l} \Big( {2 \over 3 s}  + { 3 tu + 2u^2 \over 4 s^3} + 4{ u - 2 t \over s^4} \tau^2_{2l}  
%- {2 \tau_{4l} \over s^2  }  - {4 \tau_{4l}^2  \over s^2 t }\Big)+
% \tau_{2l} \Big( {11 \over 6s} -  {3 t u + 2 u^2 \over 4 s^3} - {4 \tau_{4l}^2 \over s^2 u} \Big)  \no\\
%\! \!  &  \! \!   \! \! & \! \!  - {  \tau_{4l} \over 6 s} -  l^2 \Big( {t - 2u \over 6 s^2 } - {(t + 3 u) \tau_{1l} + ( u - t) \tau_{2l}\over s^3} \Big) 
%+ 8 i {(\tau_{1l}+ \tau_{2l})^2 \over s^4} \epsilon (k_1,l_2,k_4, l) \ . \no \\
\eea
All the other numerators can be obtained from the ones shown here. Specifically:
\begin{itemize}
 \item the numerators of the third box can be obtained using the Bose symmetry (\ref{exchange});
 \item the expressions for the numerators $\hat n_{i,r^1_1},\hat n_{i,r^1_1r^2_2}$ and $\hat n_{i,r^1_1r^2_2r^3_3}$ follow from the requirement that the 
 amplitudes reproduce the known formulae in the $\cN=1$ and $\cN=2$ cases, as in eqs.~(\ref{condN4}-\ref{condN1}); 
  \item the numerators for the triangle and bubble graphs are obtained using the kinematic Jacobi relations \eqref{Jacobifirst}-\eqref{Jacobilast}. 
\end{itemize}
Some of the free coefficients that are not fixed by the cut conditions have been chosen such that one bubble, two triangle and all snail 
graphs are set to zero,
\be 
\hat n_{i;R_l} \equiv 0 \ , \qquad (\forall) \ R_l \ , \ \quad i=4,6,10  \quad \text{and} \quad i>12  \ .
\ee 
The remaining coefficients have been fixed to obtain a particularly simple representation of the amplitude.
We emphasize that, in the construction of the numerators we have used four-dimensional cuts and therefore some rational terms
are not accounted for. We have checked that, upon integration, the graphs with numerator factors $n_{i;1}$ yield the pure YM four-gluon 
amplitude \cite{Bern:1995db} up to such terms.
%It should be emphasized that the amplitude presentation we have obtained involves integrals in $D= 4 - 2 \epsilon$ dimensions 
%and misses rational terms. 
%We have explicitly verified that the amplitude reproduces the known result after integration. 
%
One may, alternatively, turn this around and use the numerator factors of \cite{Bern:2013yya} of the pure YM four-gluon amplitude in 
BCJ form to construct $n_{i;1}$ and with it construct the amplitudes of all non-supersymmetric orbifolds.

\subsection{Four-scalar 
%matter 
amplitudes}

Four-scalar amplitudes are the simplest amplitudes with external legs charged under the orbifold generators. We will 
focus here on the particular field configuration $\big(1^{ \phi} 2^{\phi} 3^{\bar  \phi} 4^{\bar \phi} \big) $
and construct the corresponding one-loop amplitude $ {\cal A}^{1-\text{loop}}_{4} \big(1^{ \phi} 2^{\phi} 3^{\bar \phi} 4^{\bar \phi} \big)$
for $\cN=2$, $\cN= 1$ and $\cN=0$ orbifolds. An important simplification compared to gluon amplitudes comes from the expectation that
these amplitudes have manifestly local numerator factors; this is  a consequence of the observation \cite{Bern:2013yya}
that the non-locality of the numerator factors of one-loop four-gluon amplitudes in pure $\cN=2$ and $\cN=1$ sYM theories
is a consequence of the use of helicity states for the external fields.

\subsubsection{Four-scalar amplitudes with $\cN=2$ supersymmetry}

As in the case of gluon amplitudes with $\cN=2$ supersymmetry (see section~\ref{glueNeq2}) , without loss of generality 
we take the $SU(4)$ orbifold matrices matrices $r$ to be as in eq.~\eqref{Neq2_r}
\be
r={\rm diag}(1,1,r_3^3,r_4^4) \ .
\ee
We focus on the amplitude $ {\cal A}^{1-\text{loop}}_{4} \big(1^{ \phi^{13}} 2^{\phi^{13}} 3^{ \phi^{24}} 4^{\phi^{24}} \big)$. The phase factors associated to the external legs are then,
\be 
R_{1}=R_{2}= r^3_3, \qquad R_{3}=R_{4}=r^4_4={1 \over r^3_3}  \ ,
\ee
while the set of phase factors capturing the action of the orbifold group on (the $SU(4)$ representation of) the internal line carrying the loop 
momentum $l$ is given by,
\be 
R_l\in{\cal R} = \big\{ 1 , r^3_3, r^4_4 \big\} \ .  
\ee

Unlike the gluon amplitude, constraints on the numerator factors are less severe here; in particular, since the external  states are changed 
under the orbifold group, the $S_2$ symmetry permuting $r_3^3$ and $r_4^4$ cannot be a symmetry of this amplitude. However, the requirement 
that as $\Gamma={\bf 1}$ the amplitude reduces to that of the $\cN=4$ theory relates the numerator factors as
\bea
n_{j;1} + n_{j;r^3_3}+ n_{j;r^4_4} &=& n_{j}^{\cN=4} \equiv i s^2 ,\qquad j=1,2,3
\cr
n_{j;1} + n_{j;r^3_3}+ n_{j;r^4_4} &=& 0 \qquad \qquad \qquad ,\qquad j\ge 4 \ .
\label{recoverNeq4_4s}
\eea
Together with the Jacobi relations \eqref{Jacobifirst}-\eqref{Jacobilast}, these equations imply that six of the nine box integrals can be 
chosen as master graphs.  

For their numerator factors we use a manifestly local ansatz
\be 
n_{j,R_l} (l) = i \hat n_{j,R_l} (l) = i {\cal P}_{2} \big( s, t \big)
\ee
with a different degree-two polynomials ${\cal P}_{2}$ for each graph and phase factor $R_l$. The result of the unitarity cut calculation 
is sufficiently simple for us to list explicitly  all numerator factors: 
%to allow us to list explicitly all numerator factors: 
\bea
\begin{array}{|c||c|c|c|c|c|c|c|c|c|c|c|c|}
\hline
%\text{\tiny{\backslashbox{\raise3ex\hbox{\hskip2ex{\large{${R_l}$}}}}{\lower3ex\hbox{\large{${\hat n}_i$}}}}}
%R_l\backslash {\hat n}_i
\text{\diaghead{\theadfont  XXXXXX}{\lower.25em\hbox{$R_l$}}{\raise.25em\hbox{${\hat n}_i$}}}
&\hat n_{1} & \hat n_2 & \hat n_3 & \hat n_4 & \hat n_5 & \hat n_6 & \hat n_7 & \hat n_8 & \hat n_9 & \hat n_{10} 
& \hat n_{11} & \hat n_{12} \cr
%R_l\backslash&\hat n_{1,R_l} & \hat n_{2,R_l} & \hat n_{3,R_l} & \hat n_{4,R_l} & \hat n_{5,R_l} & \hat n_{6,R_l} & \hat n_{7,R_l} 
%& \hat n_{8,R_l} & \hat n_{9,R_l} & \hat n_{10,R_l} & \hat n_{11,R_l} & \hat n_{12,R_l} \cr
\hline
\hline
       1& s^2& -st& 0 &    0 &st &  0 & st  &  su &  -su&   0 & -2su& -2st  \cr
\hline       
r_3^3 & 0&   0 &   0&     0&  0&  0 &  -st &    0 &  su& 0&  su  &  st   \cr
\hline
r_4^4& 0&   -su & s^2& 0& -st & 0 & 0 & -su&   0&      0&   su &  st   \cr
\hline
\end{array} \ .
\eea
\iffalse
\bea 
\hat n_{1,1} = s^2 \ , && \\
 \hat n_{2,1} = -st \ , &&  \hat n_{2,r^4_4} = - su  \\
 \hat n_{3,r^4_4} = s^2  \ , &&  \\
  \hat n_{5,1} = st \ ,& \qquad&  \hat n_{5,r^4_4} = - st \ ,  \\
  \hat n_{7,1} = st \ , & \qquad & \hat n_{7,r^3_3} = - st \ , \\
  \hat n_{8,1} = su \ , & \qquad &  \hat n_{8,r^4_4} = - su \ , \\
  \hat n_{9,1} = su \ , & \qquad & \hat n_{9,r^3_3} = - su \ ,\\
  \hat n_{11,1} = -2su \ , & \qquad & \hat n_{11,r^3_3} =  su \ , \qquad \quad  \hat n_{11,r^4_4} = su \ ,  \\
  \hat n_{12,1} = -2st \ , &\qquad & \hat n_{12,r^3_3} =  st \ , \qquad \quad  \hat n_{12,r^4_4} = st \ . 
    \eea
 \fi   
Each entry is the numerator factor of the graph specified by the top entry of the column dressed with the orbifold phase 
specified by the left-most entry of the row. We note that the sum of the entries of each column gives the numerator factor 
of that graph in the $\cN=4$ sYM four-scalar amplitude, cf. eq.~\eqref{recoverNeq4_4s}.
We also note that, perhaps due to the large amount of supersymmetry, the Levi-Civita tensor is absent from all numerator factors.

\subsubsection{Four-scalars amplitudes with $\cN=1$ supersymmetry}

The four-scalar amplitude in $\cN=1$ supersymmetric orbifold theories are very interesting as they exhibit some of the features 
of non-supersymmetric theories while still being relatively compact. As in the case of the gluon amplitude in these theories we choose 
the $r$ matrix representing the orbifold group inside $SU(4)$ as
\be
r={\rm diag}(1, r_2^2, r_3^3, r_4^4) ~,\qquad r_2^2 r_3^3 r_4^4 = 1 \ .
\ee
We focus on the amplitude $ {\cal A}^{1-\text{loop}}_{4} \big(1^{ \phi^{12}} 2^{\phi^{12}} 3^{ \phi^{34}} 4^{\phi^{34}} \big)$.
The phases associated to the external legs are then,
\be 
R_{1}=R_{2}= r^2_2, \qquad R_{3}=R_{4}=r^3_3 r^4_4={1 \over r^2_2}  \ .  
\label{external_phases_Neq1_4s}
\ee
The set of phases $R_l$ dressing the internal leg  carrying the loop momentum $l$ is 
\be 
R_l\in{\cal R} = \big\{ 1 , r^2_2, r^3_3,  r^4_4, r^2_2 r^3_3, r^2_2 r^4_4, r^3_3 r^4_4 ,
(r^2_2)^2, (r^2_2)^2 r^3_3, (r^2_2)^2 r^4_4, (r^3_3)^2 r^4_4, r^3_3 (r^4_4)^2, (r^3_3)^2 (r^4_4)^2 \big\} \ . 
\ee
This set is larger than the one in eq.~\eqref{Neq1_gluon_phases} for the gluon amplitude in part due to the $SU(4)$-charge flow 
between external legs.  We should also note that, unlike the previous three examples, some of the phases above (the last six 
elements of the set) cannot be obtained from the R-symmetry labels of a \emph{physical} particle going around the loop. 
However, a solution to the kinematic Jacobi relations \eqref{Jacobifirst}-\eqref{Jacobilast} appears to exist only if these fictitious 
particles are included. The appearance of these extra representations at the intermediate steps of the computation
should not be a surprise because physical  amplitudes are obtained only after the summation over all the orbifold elements $(r,g)$.
The two-particle cuts of such an ansatz can be correct order by order in the $r^a_a$  
only if all cuts containing at least one unphysical particle vanish. This requires that 
the numerator factor of an unphysical  graph contains an inverse propagator for at least 
one unphysical particle in all cuts. 

From the $S_3$ symmetry permuting the  the nontrivial elements of $r$, the $S_2$ subgroup interchanging $r_3^3$ and $r_4^4$
preserves the external line phase factors \eqref{external_phases_Neq1_4s} and thus can be used to impose the following relations:
\be 
n_{i; r^4_4} = n_{i;r^3_3}  \ , \quad  n_{i; r^2_2 r^4_4} = n_{i; r^2_2 r^3_3} \ , \quad 
 n_{i; (r^2_2)^2 r^4_4} = n_{i; (r^2_2)^2 r^3_3} \ , \quad  n_{i; r^3_3 (r^4_4)^2} = n_{i; (r^3_3)^2 r^4_4} \ .
\ee
Of the maximum of 39 box master integrals we are therefore left with 27.  Last but not least, further numerator relations come from the
requirement that as $\Gamma={\bf 1}$ the amplitude reduces to that of $\cN=4$ sYM theory. We will not write them out explicitly. 

As in the $\cN=2$ four-scalar amplitude example, the numerator factors are expected to be local so we use the ansatz 
\be 
n_{i,R_l} (l) = i \hat n_{i,R_l} (l) = i {\cal P}_{2;1}\big( s, t, \tau_{1l} ,\tau_{2l}, \tau_{3l}, l^2 \big) 
- c \ \epsilon (k_1,k_2,k_4,l) \ , 
\ee
where $c$ is a real constant and ${\cal P}_{2;1}$ is a degree-two polynomial which also has at most unit degree in its $l$-dependent arguments ($c$ 
and ${\cal P}_{2;1}$ are different for different graphs). The unitarity cuts determine the numerator factors of the master graphs:
%. We list 
%only the non-zero independent numerator factors:

\noindent $\bullet$ The graph topology (1) in fig.~\ref{intbox}
%%%%%%%%%%%%%%%%%%%%%%%%%
\iffalse
\bea 
\hat n_{1;1} & =  & s^2 - {s \over 3}  (\tau_{1l} - \tau_{2l} + l^2) \ , \label{nn-n1s-first} \\
\hat n_{1;r^2_2} & = & - {s \over 12}  (5 \tau_{1l} + \tau_{2l} + 2 l^2) \ ,  \\
\hat n_{1;r^3_3} & = & - {s \over 12}  (\tau_{1l} + 5 \tau_{2l} - 2 l^2) \ ,  \\
\hat n_{1;r^2_2r^3_3} & = &  {s \over 12}  (5 \tau_{1l} + \tau_{2l} + 2 l^2) \ , \\
\hat n_{1;r^3_3r^4_4} & = & {s \over 12}  (\tau_{1l} + 5 \tau_{2l} - 2 l^2) \ .  
\eea
\fi
%%%%%%%%%%%%%%%%%%%%%%%%%
\bea 
\hat n_{1;1} & =  & s^2 - {s \over 3}  (\tau_{1l} - \tau_{2l} + l^2) \ , \label{nn-n1s-first}  \\
\hat n_{1;r^2_2}  &=&  - {s \over 12}  (5 \tau_{1l} + \tau_{2l} + 2 l^2) \quad \ , \quad~~
\hat n_{1;r^3_3}  = - {s \over 12}  (\tau_{1l} + 5 \tau_{2l} - 2 l^2) \ ,  \\
\hat n_{1;r^2_2r^3_3} & = &  {s \over 12}  (5 \tau_{1l} + \tau_{2l} + 2 l^2) \quad ~~\ , \quad
\hat n_{1;r^3_3r^4_4}  =  {s \over 12}  (\tau_{1l} + 5 \tau_{2l} - 2 l^2) \ .  
\eea

\noindent $\bullet$ The graph topology (2) in fig.~\ref{intbox}
\bea 
\hat n_{2;1} \! \! & = & \! \!  {5 \over 24} su - st -  (\tau_{1l} + \tau_{2l} )\Big( t + {2\over 3} s\Big) + {s \over 4}  (\tau_{4l} + \tau_{2l} - l^2)  - 2 i \epsilon(k_1,k_2,k_4,l) \ ,
\qquad \\
\hat n_{2;r^2_2} \! \! & = & \! \! - {s \over 12}  ( \tau_{1l} - \tau_{4l} +  l^2) \ ,  \\
\hat n_{2;r^3_3} \! \! & = & \! \!  -{5 \over 24} su   +   (\tau_{1l} + \tau_{2l} ) \Big( t + {7 \over 12} s \Big) - {s \over 6}  (\tau_{4l} + \tau_{2l} - l^2)  
+ 2 i \epsilon(k_1,k_2,k_4,l)  ,\qquad \\
\hat n_{2;r^2_2r^3_3} \! \! & = & \! \!  {s \over 12}  ( \tau_{1l} - \tau_{4l} +  l^2) \ ,  \\
\hat n_{2;r^3_3 r^4_4} \! \! & = & \! \! - {5 \over 6} su -  (\tau_{1l} + \tau_{2l} ) \Big( t + {7 \over 12} s \Big) + {s \over 4}  (\tau_{4l} + \tau_{2l} - l^2)  
- 2 i \epsilon(k_1,k_2,k_4,l) \ , \\
\hat n_{2;(r^3_3)^2r^4_4} \! \! & = &  \! \!  {s \over 24}  ( u - 2 \tau_{2l} - 2 \tau_{4l} + 2 l^2) \ ,  \\
\hat n_{2;(r^3_3 r^4_4)^2} \! \! & = & \! \! -{s \over 24}  ( u - 2 \tau_{2l} - 2 \tau_{4l} + 2  l^2) \ . 
\label{nn-n1s-last} \eea

\noindent $\bullet$ The graph topology (3) in fig.~\ref{intbox} can be obtained employing the Bose symmetry for the exchange of particles $1$ and $2$.
%\bea \hat n_{3;1} \!\!& = &\!\!  - {5 \over 24} s^2 - {s \over 12} (  5 \tau_{1l} + 5 \tau_{2l} + 2 l^2) \ ,  \\
%\hat n_{3;r^3_3} \!\!& = &\!\!  {5 \over 24} s^2 + {s \over 12} (  5 \tau_{1l} + 5 \tau_{2l} + 2 l^2)  ~ \ , ~
%\hat n_{3;r^3_3r^4_4} =  {5 \over 6} s^2 - {s \over 3} (   \tau_{1l} +  \tau_{2l} +  l^2) \ , \\
%\hat n_{3;(r^3_3)^2r^4_4} \!\!& = &\!\! - {s \over 24}   ( s +2 \tau_{1l} + 2 \tau_{2l} - 4 l^2)  ~ \ , ~
%\hat n_{3;(r^3_3)^2(r^4_4)^2} =   {s \over 24}  ( s +2 \tau_{1l} + 2 \tau_{2l} - 4 l^2)  \ . ~~~~
%\label{nn-n1s-last}
%\eea

\noindent
We note that, as required, the numerator factors corresponding to graphs with some unphysical fields going around the loop 
can be expressed as a linear combination of inverse propagators so that the corresponding integrals 
contribute to only one of the three two-particles cuts -- the one that cuts only physical internal lines.
As an example of this, we consider the numerator factor  $\hat n_{1; r^3_3 r^4_4}$. 
It is easy to see that the  internal line between the momenta $k_2$ and $k_3$ may be unphysical (with phase $(r^3_3 r^4_4)^2$). 
The numerator factor can be written as
\be
\hat n_{1, r^3_3 r^4_4} = {s \over 24} \big( (l + k_1)^2 - 5 (l - k_2)^2 \big) \ .
\ee
The second inverse propagator removes the propagator corresponding to the unphysical particle, 
while the first inverse propagator appears to be problematic. 
However, inspecting the numerator factors of the second and seventh graph,
\be
\hat n_{2; (r^3_3 r^4_4)^2} = - {s \over 24} \big( (l - k_2 - k_4)^2 + l^2 \big) \ , \quad
\hat n_{7; (r^2_2)^2} = - {s \over 24} \big( t - (l - k_2)^2 - (l + k_3)^2 + 4 l^2 \big) \ , 
\ee
we note that the first term of the first graph, the second term of the second graph and the first term of the seventh graph 
are all proportional to the same triangle integral $I_3(t)$; the proportionality constant includes the combination of color factors, 
\be 
c_{1; r^3_3 r^4_4} - c_{2; (r^3_3 r^4_4)^2 } - c_{7; (r^2_2)^2}
\ee
which vanishes due to one of the color Jacobi identities. 
For all other numerator factors, one can see that in the box color structures
all terms in which the propagator of an unphysical state is not removed vanish due to similar cancellations.

The triangle color structures leave behind terms with unphysical fields going around the loop in a ``snail'' integral (where the bubble is attached to the vertex). 
Similarly, the bubble color structures leave behind some tadpole integrals (where the bubble is attached to the internal propagator). 
Such snail and tadpole integrals are not constrained by the standard two-particles cuts we have employed; however, 
they vanish upon integration for massless external particles in $D=4-2\epsilon$ dimensions.

It is also easy to verify that, by setting $r_2^2=r_3^3=r_4^4=1$ and summing all the numerators corresponding to the same labeled 
graph we obtain the standard numerator factors of the one-loop four-scalar amplitude $\cN =4$ sYM theory.

As in the previous examples we have imposed Bose symmetry for the exchange of external particles $1$ and $2$ (as well as $3$ and $4$);
it requires that
\be 
\hat{n}_{1,R_l}(l) = \hat{n}_{3,R_l/R_1}(l-k_1)\big|_{k_1 \leftrightarrow k_2} , \qquad 
\hat{n}_{2,R_l}(l) = \hat{n}_{2,R_4/R_l}(k_4-l)\big|_{k_1 \leftrightarrow k_2} 
\label{exchangescalars} \ .
\ee
Moreover, we have required that the numerator factors of snail integrals obey the relation (\ref{redsnail}) and 
we have set to zero the numerators of one bubble and two triangle graphs,
\be 
\hat n_{i;R_l} \equiv 0 \ , \qquad (\forall) \ R_l \ , \ \quad i=4,6,10  . 
\ee 
Finally, we have required that upon setting $r_2^2=1$ and summing over the numerator factors corresponding to identical graphs
we reproduce $\cN=2$ orbifold amplitude derived in the previous section. These conditions fix all coefficients of the ansatz.

The expressions (\ref{nn-n1s-first}-\ref{nn-n1s-last}) solve all the Jacobi-like relations (\ref{Jacobifirst}-\ref{Jacobilast}). However, 
the conditions for having vanishing tadpole numerators  (\ref{Jacobitad}) are not generically satisfied. 
It is very interesting to note that the color factor of a tadpole graph always contains a term of the form $f^{ABC}g^{AB}$, 
which can be expressed in the trace basis as,
\be f^{ABC}g^{AB} = 2 i \ \Im \Big( \text{Tr} g \  \text{Tr} g^{\dagger} T^C  \Big) \ . \label{tadcanc} \ee
Thus, the right-hand side vanishes for any regular orbifold due to (\ref{regular}), implying that amplitudes in such theories 
have no tadpole graphs\footnote{The converse is not necessarily true, and it might be possible to find examples of non-regular 
orbifolds with vanishing tadpole integrands.}. This mirrors the string theory result, where regularity of the orbifold guarantees tadpole 
cancellation.

%%%%%%%%%%%%%%%%%%%%%%%%%%%%%%%%%%%%%%%%%%
\subsubsection{Four-scalars amplitudes with no supersymmetry}

The last example we discuss is the one-loop four-scalar amplitude $ {\cal A}^{1-\text{loop}}_{4} \big(1^{\phi^{12}}2^{ \phi^{12}}3^{ \phi^{34}}4^{ \phi^{34} }\big)$
in non-supersymmetric orbifold theories. The $SU(4)$ matrix $r$ generating the orbifold group is now a general diagonal matrix of phases subject 
to the unit determinant condition:
\be
r={\rm diag}(r_1^1, r_2^2, r_3^3, r_4^4) \ ,\quad \det(r) = 1 \ .
\ee
The external line phases for the amplitude $ {\cal A}_{4} \big( 1^{\phi^{12}}2^{ \phi^{12}}3^{ \phi^{34}}4^{ \phi^{34}} \big)$ 
are
\be 
R_{1}=R_{2}= r^1_1 r^2_2, \qquad R_{3}=R_{4}=r^3_3 r^4_4={1 \over r^1_1 r^2_2}  \ ,
\label{external_phases_Neq0}
\ee
while the internal line phases $R_l$ belong to the 33-element set 
\bea R_l\in{\cal R} \!\!\!&=&\!\!\! \big\{ 1 , r^1_1, r^2_2, r^3_3,  r^4_4, r^1_1 r^2_2, 
r^1_1 r^3_3, r^1_1 r^4_4, r^2_2 r^3_3, r^2_2 r^4_4, r^3_3 r^4_4 , r^1_1r^2_2 r^3_3 , r^1_1r^2_2 r^4_4 , r^1_1r^3_3 r^4_4 , r^2_2r^3_3 r^4_4,\no \\
&& (r^1_1)^2r^2_2, r^1_1(r^2_2)^2, (r^3_3)^2r^4_4, r^3_3(r^4_4)^2, (r^1_1r^2_2)^2, (r^1_1)^2 r^2_2 r^3_3, (r^1_1)^2 r^2_2 r^4_4, r^1_1 (r^2_2)^2 r^3_3, \no \\
&&r^1_1 (r^2_2)^2 r^4_4,  
r^1_1 (r^3_3)^2 r^4_4, r^2_2 (r^3_3)^2 r^4_4, r^1_1 r^3_3 (r^4_4)^2 ,
 r^2_2 r^3_3 (r^4_4)^2, (r^3_3 r^4_4)^2,(r^1_1 r^2_2)^2 r^3_3, \no \\ && (r^1_1 r^2_2)^2 r^4_4, r^1_1( r^3_3r^4_4)^2 , r^2_2( r^3_3r^4_4)^2 \big\} \ . 
\eea
As  explained in the case of the same amplitude in $\cN=1$ orbifold theories, while some of the elements of ${\cal R}$  cannot be obtained 
from the R-symmetry labels of a physical particle going around the loop, they appeared necessary for the existence of a solution of the 
kinematic Jacobi relations which is consistent with all unitarity cuts. 
The external phases \eqref{external_phases_Neq0} are invariant under the $S_2\times S_2\subset S_4\subset SU(4)$ symmetry permuting 
$r_1^1$ and $r_2^2$ and, independently, $r_3^3$ and $r_4^4$. This symmetry implies that the 99 numerator factors of box integrals are related as
\bea 
&& n_{i; r^2_2} = n_{i;r^1_1} \ , \qquad  n_{i; r^4_4} = n_{i;r^3_3}  \ , \quad  n_{i; r^2_2 r^4_4} = n_{i; r^2_2 r^3_3} = n_{i; r^1_1 r^4_4}=  n_{i; r^1_1 r^3_3} \ ,
\no \\ && n_{i; r^1_1 r^2_2 r^4_4} = n_{i;r^1_1 r^2_2 r^3_3} \ , \quad n_{i; r^2_2 r^3_3 r^4_4} = n_{i;r^1_1 r^3_3 r^4_4} \ ,
\quad n_{i; r^1_1 (r^2_2)^2} = n_{i; (r^1_1)^2 r^2_2} \ , \no \\ &&  
n_{i; r^3_3 (r^4_4)^2} = n_{i; (r^3_3)^2 r^4_4} \ , \quad n_{i; r^1_1 (r^2_2)^2 r^3_3} = n_{i; r^1_1 (r^2_2)^2 r^4_4} = n_{i; (r^1_1)^2 r^2_2 r^3_3} =
n_{i; (r^1_1)^2 r^2_2 r^4_4}\ , \no \\
&& n_{i; r^1_1 r^3_3 (r^4_4)^2} = n_{i; r^2_2 r^3_3 (r^4_4)^2}=n_{i; r^2_2 (r^3_3)^2 r^4_4} = n_{i; r^1_1 (r^3_3)^2 r^4_4} \ , \quad  
n_{i; r^2_2 (r^3_3 r^4_4)^2} = n_{i; r^1_1 (r^3_3 r^4_4)^2} \ , \no \\ &&   n_{i; (r^1_1r^2_2)^2 r^4_4} = n_{i; (r^1_1r^2_2)^2 r^3_3} \ .   
\eea
We are therefore left with 48 master integrals (the three additional relations, which we do not write explicitly, imposing that as $r_i^i=1$ 
we recover the $\cN=4$ sYM amplitude reduce this number down to 45).

As for the other scalar amplitude examples, the ansatz for the numerator factors of the master integrals is manifestly local:
\be 
n_{i,R_l} (l) = i \hat n_{i,R_l} (l) = i {\cal P}_{2;2}\big( s, t, \tau_{1l} ,\tau_{2l}, \tau_{3l}, l^2 \big)s 
- c \ \epsilon (k_1,k_2,k_4,l) \ ; 
\ee
here, as before, $c$ is a real constant while ${\cal P}_{2;2}$ is a polynomial (different for each graph) of degree two which is also up to degree two in its 
loop-momentum-dependent arguments. The unitarity cuts and the additional conditions explained below determine the numerator factors to be (we 
list only the non-vanishing ones):

\noindent $\bullet$ The graph topology (1) in fig.~\ref{intbox}
\bea \hat n_{1;1} & \! \! =  & \! \! s^2 + (\tau_{1l} - \tau_{2l}) \Big(  \tau_{1l} - \tau_{2l} + {4 \over 3} l^2 \Big) + {4 \over 3} \tau_{1l} \tau_{2l} + {2 \over 3} l^4  \ , \label{nn-n0s-first} \\
\hat n_{1;r^1_1} & \! \! = & \! \! {1 \over 6}   (\tau_{1l} + \tau_{2l}) \big( {s } + \tau_{2l} + \tau_{1l} \big) - {4 \over 3} \tau_{1l} \big(\tau_{1l} + l^2 \big)
+ {s \over 6} \big(2 \tau_{2l}- l^2 \big) - {l^4 \over 3}  \ ,  \\
\hat n_{1;r^3_3} & \! \! = & \! \! - s   \tau_{2l} -   (\tau_{1l} + \tau_{2l}) \Big( {t \over 2} + \tau_{4l} \Big) + {4 \over 3} \tau_{1l} \tau_{2l}
+l^2 \Big( {s\over 2} + 2 {\tau_{2l}-\tau_{1l} \over 3} \Big)- {l^4 \over 3}  \ ,  \\
\hat n_{1;r^1_1r^2_2} & \! \! = & \! \!  -  (\tau_{1l} + \tau_{2l}) \Big( {7 \over 12}s + {\tau_{2l} + \tau_{1l} \over 6}  \Big) +
{4 \over 3}\tau_{1l} \big(\tau_{1l} + l^2  \big)  + {l^4 \over 3}  \ , \\
\hat n_{1;r^1_1r^3_3} & \! \! = & \! \!  \tau_{4l} (\tau_{1l} + \tau_{2l})+ (\tau_{1l} -\tau_{2l}) \Big( {2 \over 3}  l^2 - {s \over 12}  \Big) 
- {u \over 2} \tau_{2l} +  {t \over 2} \tau_{1l}
-{4 \over 3} \tau_{1l} \tau_{2l} - l^2{s - l^2 \over 3}    , \qquad \\
\hat n_{1;r^3_3r^4_4} & \! \! = & \! \! - {1 \over 6} (\tau_{1l} + \tau_{2l})^2+ {4\over 3} \tau_{2l} \big(  \tau_{2l} -  l^2   \big) 
+ {7 \over 12} s \big( \tau_{1l} + \tau_{2l} \big)   + {l^4 \over 3}  \ , \\
\hat n_{1;r^1_1 r^2_2 r^3_3} & \! \! = & \! \! - (\tau_{1l} + \tau_{2l}) \Big( \tau_{4l} + {t \over 2} \Big) - (\tau_{1l} -\tau_{2l}) 
\Big( {2 \over 3}  l^2 - {s \over 2}  \Big)
+{4 \over 3} \tau_{1l} \tau_{2l} + {s \over 2} l^2  - {l^4 \over 3}  \ , \\  
\hat n_{1;r^1_1r^3_3 r^4_4} & \! \! = & \! \! {1\over 6}(\tau_{1l} + \tau_{2l}) \big( \tau_{1l} + \tau_{2l} - {s } \big) - {s\over 3} \Big( \tau_{1l} + {l^2 \over 2} \Big) 
-{4\over 3}\tau_{2l} \big( \tau_{2l} -   l^2  \big) 
- {l^4 \over 3}  \ .
\eea

\noindent $\bullet$ The graph topology (2) in fig.~\ref{intbox}
\bea \hat n_{2;1} \! \! & = & \! \!  - st +  (\tau_{1l} + \tau_{2l} ) \Big( {7 \over 12} s - {17\over 12} t  + \tau_{1l} + {7 \over 6} \tau_{2l} +{\tau_{4l}\over 6}\Big)
+ 2 s  (\tau_{4l} - \tau_{1l}) \  \no \\ 
\! \! &  & \! \! \qquad - {4 \over 3 } \tau_{1l} \tau_{4l} - l^2 \Big( 2 s - {u \over 6} + {\tau_{2l}-\tau_{1l} \over 2} + \tau_{4l} \Big) + {l^4 \over 2} \ , \qquad \\
\hat n_{2;r^1_1} \! \! & = & \! \!  {\tau_{1l} \over 6}  (3s+ \tau_{1l} + \tau_{2l} )+ {\tau_{4l} \over 6} ( 3\tau_{1l}- \tau_{2l} -3 s )
 + {l^2\over 6} \big( 3 s+  \tau_{2l}-\tau_{1l} + 2 \tau_{4l} \big) - {l^4 \over 6} , \qquad \ \\
\hat n_{2;r^3_3} \! \! & = & \! \!  - {su \over 2 }+  (\tau_{1l} + \tau_{2l} ) \Big(  t  -{\tau_{4l}\over 3}\Big)
+ {s \over 2}   (2\tau_{2l} + \tau_{4l} )- {2 \over 3} \tau_{2l} \tau_{4l}  \no \\
\! \! &  & \! \! \qquad  + l^2 \Big( {t \over 6}  - {7 \over 12} s + {\tau_{2l}-\tau_{1l} \over 3} + {2 \over 3}\tau_{4l} \Big) - {l^4 \over 3} + 2i \epsilon (k_1,k_2,k_4,l)\ ,
\quad \\
\hat n_{2;r^1_1 r^2_2} \! \! & = & \! \! {\tau_{4l} } \Big( {7 \over 12} s -
{\tau_{1l}\over 2}+{ \tau_{2l} \over 6} \Big)
-  {\tau_{1l} }  \Big({7\over 12} s + {\tau_{1l} + \tau_{2l}\over 6} \Big)- \no \\
&& \qquad {l^2} \Big( {7 \over 12}s  + {\tau_{2l}- \tau_{1l} \over 6 }+{ \tau_{4l}\over 3} \Big) + {l^4 \over 6} \ , \qquad \quad \\
\hat n_{2;r^1_1 r^3_3} \! \! & = & \! \!  s   \Big({7\over 12} \tau_{1l}- {\tau_{4l}\over 12} \Big)+ {\tau_{4l} } 
{\tau_{1l} + 3\tau_{2l} \over 3} +
  {l^2 \over 6} \big( 2 s - \tau_{1l}- 3\tau_{2l} - 2 \tau_{4l} \big) + {l^4 \over 6} \ , \qquad \\
\hat n_{2;r^3_3 r^4_4} \! \! & = & \! \!  - {5 \over 4} su +  (\tau_{1l} + \tau_{2l} ) \Big( -{3 \over 2} t +{7\over 6}\tau_{1l} +\tau_{2l} -{\tau_{4l}\over 6}\Big)
- {2 \over 3} \tau_{1l} \tau_{4l} \no \\
\! \! &  & \! \! \qquad - l^2 \Big( {t \over 3}  + {9 \over 4} s + {\tau_{2l}-\tau_{1l} \over 2} + \tau_{4l} \Big) + s \Big(  \tau_{2l} - {11\over 12} \tau_{1l}  + {23 \over 12} \tau_{4l} \Big)+
{l^4 \over 2} \ ,
\qquad \\
\hat n_{2;r^1_1 r^2_2 r^3_3} \! \! & = & \! \! - {s \over 2} {\tau_{1l} }- {\tau_{4l} }  
{\tau_{1l}+ 3 \tau_{2l} \over 3} +
  {l^2\over 6} \big( -{3\over 2} s  + \tau_{1l}+3 \tau_{2l} + 2 \tau_{4l} \big) - {l^4 \over 6} \ , \qquad \quad \eea 
  \bea
\hat n_{2;r^1_1r^3_3 r^4_4} \! \! & = & \! \!   {5 \over 24} su +  (\tau_{1l} + \tau_{2l} ) \Big( {5 \over 12} t -{7\over 6}(\tau_{1l} +\tau_{2l} )\Big)
+ {2 \over 3} \tau_{1l} \tau_{4l} - s \Big(   \tau_{2l} - { \tau_{1l} \over 4}  + {5 \over 4} \tau_{4l} \Big) \no \\
\! \! &  & \! \! \qquad + l^2 \Big( {t \over 6}  + {17 \over 12} s + {\tau_{2l}-\tau_{1l} \over 3} + {2 \over 3}\tau_{4l} \Big) -
{l^4 \over 3} -  2i \epsilon (k_1,k_2,k_4,l) \ ,
\qquad \\
\hat n_{2; (r^3_3)^2r^4_4} \! \! & = & \! \!  -{su \over 4} + {s\over 2} (\tau_{2l} }+ {\tau_{4l} )  
- {l^2\over 6} \big(  2 s - t + 3 \tau_{1l}+ \tau_{2l} - 2 \tau_{4l} \big) - {l^4 \over 6} \ , \qquad \quad \\
\hat n_{2; r^1_1(r^3_3)^2 r^4_4} \! \! & = & \! \!  {7 \over 24} su -{7\over 12}s (\tau_{2l} }+ {\tau_{4l} )  
+ {l^2\over 6} \Big(  {5\over2} s - t + 3 \tau_{1l}+ \tau_{2l} - 2 \tau_{4l} \Big) + {l^4 \over 6} \ , \qquad \quad \\
\hat n_{2; (r^3_3 r^4_4)^2} \! \! & = & \! \!  (\tau_{1l}+\tau_{2l}-3 s) \Big({u \over 12} - {\tau_{2l}+\tau_{4l}\over 6} \Big)  
- {l^2\over 6} \big(  4 s + t - \tau_{1l}+ \tau_{2l} + 2 \tau_{4l} \big) + {l^4 \over 6} \ , \qquad \quad \\
\hat n_{2; r^1_1(r^3_3 r^4_4)^2} \! \! & = & \! \!  \big({5\over 2}s-\tau_{1l}-\tau_{2l}\big) \Big({u \over 12} - {\tau_{2l}+\tau_{4l}\over 6} \Big)  
+ {l^2\over 6} \big(  {7\over 2} s + t - \tau_{1l}+ \tau_{2l} + 2 \tau_{4l} \big) - {l^4 \over 6} \ . \qquad 
\label{nn-n0s-last} \eea

As in the previous cases, we have imposed on the numerator factors the exchange symmetry between particles $1$ and $2$, which yields 
the relations (\ref{exchangescalars}).  Hence, the numerator factor of the third box can be found as
\be  \hat{n}_{3,R_l}(l) = \hat{n}_{1,R_l R_1}(l + k_1)\big|_{k_1 \leftrightarrow k_2}  \ ;
\ee
and will not be written explicitly.
It is not difficult to check that upon setting $r_i^i=1$ and summing the numerator factors of identical graphs ({\it i.e.} or the graphs that 
are in general different only because of the insertion of the orbifold group element in their color structure) one recovers the numerator 
factors of the one-loop four-scalar amplitude in $\cN=4$ sYM theory.

We have also required that all snail integrals obey the relation (\ref{redsnail}) for general $D$ as in the $\cN=1$ case and 
we have set  to zero the numerators of one bubble and two triangle graphs,
\be 
\hat n_{i;R_l} \equiv 0 \ , \qquad (\forall) \ R_l \ , \ \quad i=4,6,10  .
\ee 
Last but not least, we have required that, with the appropriate choices for the diagonal entries of the matrix $r$ we reproduce the 
numerator factors of the $\cN=2$ and $\cN=1$ one-loop four-scalar amplitudes  discussed in previous sections. 
%
%The remaining free coefficients to zero for simplicity.

As in the $\cN=1$ case, the numerator factors (\ref{nn-n0s-first})-(\ref{nn-n0s-last}) solve all the generalized Jacobi relations 
(\ref{Jacobifirst})-(\ref{Jacobilast}), but have non-vanishing tadpole numerators unless the right-hand side of eq.~(\ref{tadcanc}) is 
equal to zero, as it happens for regular orbifolds.  

\newpage

\renewcommand{\theequation}{6.\arabic{equation}}
\setcounter{equation}{0}
\section{On the double-copy construction of non-factorized gravity amplitudes%: a well-motivated conjecture 
\label{doublecopy}
}

Whenever a pair of (supersymmetric) gauge theories coupled with matter 
%(fermions and) scalar 
fields in the adjoint representations can be 
related to a gravitational theory through  Kawai-Lewellen-Tye relations \cite{KLT}, the amplitudes of the gravitational theory can be 
immediately obtained from a duality-satisfying presentation of the corresponding gauge amplitudes by replacing the color factors of  
one theory with the kinematic numerators of the second one corresponding the the same color factor  \cite{Bern:2008qj,Bern:2010yg}.
As summarized in earlier sections, this double-copy property of (super)gravity amplitudes was tested in $\cN=8$ supergravity as well 
as in supergravity theories with reduced supersymmetry and matter couplings. Many such theories can be obtained as factorized orbifolds 
of $\cN=8$ supergravity \cite{Carrasco:2012ca}.
%
%While they can also be obtained as orbifolds of $\cN=8$ supergravity, o
Other interesting theories -- such as pure $\cN=3$ and $\cN=2$ 
supergravities -- are however not factorized and thus it is not immediately clear how to construct their scattering amplitudes 
in terms of simpler gauge theories. The difficulty relates to the fact that the orbifold group acts on the two $\cN=4$ sYM copies 
making up the parent theory in a correlated fashion.

The formalism discussed in section~\ref{loop_orb} for the calculation of loop amplitudes in orbifold gauge theories suggests a possible approach 
to this problem.   
As we discussed at length, the numerator factors in eq.~\eqref{OrbifoldLoopGeneral} receive contributions from fields of the 
parent ($\cN=4$ sYM theory) in representations $R_{l_1},\dots,R_{l_L}$ of $\Gamma$ running in the $1,\dots, L$ loop (and the 
summation projects out the non-invariant components of fields once the $\Gamma$-representation of the color factor is accounted for). 
This is very much analogous to what we need to "factorize" a non-factorizable theory. We can therefore formulate our proposal.

We consider an orbifold supergravity theory with an abelian orbifold group $\Gamma\in SU(4)\times SU(4) \subset SU(8)$
%; we further assume that $\Gamma$ is such that  its action on the fundamental representation of $SU(8)$ is given by a matrix 
%whose upper-left and lower-right $4\times 4$ blocks generate subgroups of $SU(4)$; 
and denote by $\Gamma_1$ and $\Gamma_2$ the subgroups of $\Gamma$ in the two $SU(4)$ factors. 
At least one of them is isomorphic to $\Gamma$, while the other is at least a subgroup of $\Gamma$. 
Assuming that amplitudes of the $\Gamma_1$ and $\Gamma_2$ orbifolds of $\cN=4$ sYM are known in a color/kinematics-satisfying 
representation of the form \eqref{OrbifoldLoopGeneral} with numerator factors $n_{i; R_{l_1},\dots,R_{l_L}}$ and 
${\tilde n}_{i; R'_{l_1},\dots,R'_{l_L}}$ respectively, we expect that, for any number of external legs, the $L$-loop 
amplitudes of the $\Gamma$-orbifold of $\cN=8$ supergravity are given by
\be  
{\cal M}^{(L)} = 
\int \prod_{k=1}^L\frac{d^dl_k}{(2\pi)^d}\, %{1\over |\Gamma|}\sum_{r_k\in\Gamma}\;
\sum_{\stackrel{\scriptstyle{(R_{l_k},R'_{l_k}) \in ({\cal R}_1, {\cal R}_2)}}{\scriptstyle{R_{l_k}R'_{l_k}=1}}}\;\sum_{i \in {\cal G}_3 } \frac{1}{S_i}{ n_{i; R_{l_1},\dots,R_{l_L}} {\tilde n}_{i; R'_{l_1},\dots,R'_{l_L}} \over \prod_m p_{m,i}^2} \ . 
\label{NonfactorizedGeneral}
\ee
Here ${\cal R}_1$ and ${\cal R}_2$ are the sets of representations of $\Gamma_1$ and $\Gamma_2$ on the fields of $\cN=4$ sYM theory. The
assignment of representations to the two kinematic numerators guarantees that the supergravity fields 
that contribute to a numerator factor -- realized as the tensor product of 
the fields of the two gauge theories --  are neutral under $\Gamma$; this is realized either  
as the tensor product of invariant fields, $R_l=1=R'_l$, or as the tensor product of fields with conjugate $\Gamma$-representations, $R'_l = R_l^*$.
%
%Since the orbifold group is assumed to be Abelian, one can restrict the sum over the elements of $\Gamma$ to a single nontrivial element
%while also eliminating the division by $|\Gamma|$.

Using the fact that at one loop each field contributes independently to amplitudes both in $\cN=4$ sYM and in $\cN=8$ supergravity, one can 
easily convince oneself that this proposal is manifestly true at this order. \footnote{Matter amplitudes are generically divergent in 
supergravity theories; their representation obtained by using eq.~\eqref{NonfactorizedGeneral} with ${\cN}\le 1$ orbifold factors 
will contain tadpole integrals (which integrate to zero in dimensional regularization).}
To nonetheless illustrate it let us look at a $\Gamma={\bf Z}_2$ orbifold which acts on the fundamental representation of 
$SU(8)$ as $({\bf 1}_2, -{\bf 1}_2, {\bf 1}_2, -{\bf 1}_2)$; for this choice of $\Gamma$ 
its two parts $\Gamma_1$ and $\Gamma_2$ are $\Gamma_1=\Gamma_2={\bf Z}_2$. 
The two sets of $SU(4)$ representations are
\be
{\cal R}_1=\{1, r_3^3, r_4^4\} = \{1, -1, -1\} = {\cal R}_2 \ . 
\ee
Then, since $r_i^ir_j^j = 1$ for all choices of $i,j=3,4$ the amplitude is
\be  
{\cal M}^{(1)} = 
\int \frac{d^dl}{(2\pi)^d}\,
\sum_{i \in {\cal G}_3 } \frac{1}{S_i}{ n_{i; 1} {\tilde n}_{i; 1} + \sum_{p,q=3,4}n_{i; r_p^p} {\tilde n}_{i; r_q^q} \over \prod_m p_{m,i}^2} \ . 
\label{NonfactorizedEG}
\ee
It is not difficult to see that the second term in the numerator represents the contribution to the amplitude of four $\cN=4$ vector multiplets ({\it e.g.}
by noticing that the tensor product of two $\cN=2$ hypermultiplets yields four $\cN=4$ vector multiplets). Using also the fact that the first term, 
$n_{i; 1} {\tilde n}_{i; 1}$, yields the amplitude of $\cN=4$ supergravity coupled to two vector multiplets \cite{Tourkine:2012vx, Carrasco:2012ca},
it follows that  \eqref{NonfactorizedEG} is the four-graviton amplitude of $\cN=4$ supergravity coupled to six vector multiplets.

It is not difficult to see that this is indeed the correct result. From the perspective of the states of $\cN=8$ supergravity 
one can interpret the ${\bf Z}_2$ orbifold acting as 
 $({\bf 1}_2, -{\bf 1}_2, {\bf 1}_2, -{\bf 1}_2)$ as a ${\bf Z}_2$ 
orbifold acting as $({\bf 1}_4, -{\bf 1}_4)$; in this formulation the theory is factorized \cite{Carrasco:2012ca} and described as the double-copy of 
$\cN=4$ sYM and pure YM coupled to six real scalars -- which is precisely $\cN=4$ supergravity coupled to six vector multiplets.

At higher loops the states propagating in each loop follow a similar pattern, as eq.~\eqref{NonfactorizedGeneral} retains in each 
of them all the supergravity fields that are invariant under $\Gamma$. Thus, we expect that beyond one loop eq.~\eqref{NonfactorizedGeneral} 
holds as long as the double-copy construction \cite{BCJLoop} holds for $\cN=8$ supergravity. 

\iffalse
\draftnote{Comment on orbifolding the duality U(1); try to construct the spectrum of the simplest theory -- $Z_3 x Z_3$ one Z3 
preserving N=2 and one preserving N=1}
%
The double-copy construction exposes an additional $U(1)$ symmetry, which promotes $SU(4)\times SU(4)$ to $SU(8)$. The charges given by
\be
{\rm q}_{{}_{U(1)}}(\Phi\otimes{\tilde\Phi}) = 2q\big(h({\tilde\Phi})-h(\Phi)\big)
\ee
\fi

We note here that, while the construction described here accommodates a large class of supergravity theories with matter, it remains
difficult to construct the scattering amplitudes of pure $\cN\le 3$ supergravities. To this end it seems necessary to enhance this
double-copy construction with additional projections eliminating matter multiplets that appear together with the supergravity multiplet in 
the tensor product of $\cN\le 2$ vector multiplets.

\newpage

\renewcommand{\theequation}{7.\arabic{equation}}
\setcounter{equation}{0}
\section{Conclusions and further comments}

In this paper we discussed in detail the color/kinematics duality for general abelian orbifolds of
$\cN=4$ sYM theory with an unitary gauge group.
%%%
% %%and the extension of the resulting amplitudes to 
%%%more general quiver gauge theories which have the orbifold theory as a special point in their space of couplings.
%%%
%
%In this paper we discussed color/kinematics duality for quiver gauge theories which can be described as general orbifolds of
%$\cN=4$ sYM theory with an unitary gauge group. 
%
Such theories have matter fields in the adjoint and bi-fundamental representations of 
the gauge group; among them, corresponding to an unfaithful representation of the orbifold group in $SU(N)$ are  the 
pure ${\cal N}=1$ and ${\cal N}=2$ sYM theories as well as pure YM theory with 0, 2, 4, or 6 scalar fields, whose one-loop amplitudes in a
color/kinematics-satisfying representation were discussed previously in \cite{Carrasco:2012ca} and \cite{Bern:2013yya}, 
respectively\footnote{We note here again that, in general, non-supersymmetric orbifolds are on the "natural line", where the quartic scalar 
coupling is the same as the gauge coupling. However, such theories are not renormalizable already at one-loop unless the couplings are allowed to have different values. 
This holds, in particular, 
for the dimensional reduction of $D$-dimensional pure YM theory to four dimensions which, for $D\le 10$ can be interpreted as an 
orbifold of $\cN=4$ sYM theory, as discussed here. Inspecting the four-vector amplitudes in $\cN=4$ supergravity with matter 
computed in  \cite{Carrasco:2012ca} it is easy to see that their divergence originates form a one-loop divergence with box-graph color 
structure in the matter-coupled pure YM theory factor.}.
An interesting result is that the one-loop four-gluon amplitudes of $\cN=2$ orbifold  theories are determined by the 
$\cN=4$ and pure $\cN=2$ four-gluon one-loop amplitudes and the kinematic Jacobi relations.
More generally, the integrands of amplitudes in all orbifolds with fixed amount of supersymmetry are described by a finite 
number of polynomials in external and loop momenta; differences between theories are encoded only in their color factors.

We have also carried out a comprehensive search for field theories with only massless fields in the adjoint representation and antisymmetric 
structure constant couplings whose amplitudes can exhibit color/kinematics duality for all external states. 
%%
%We found that, with one exception which requires an inspection of a five-point amplitude, the four-point amplitudes already identify all 
%allowed interactions and their relative strengths. 
%%
%While tree-level four-point amplitudes with at least two gluons generically obey color/kinematics duality independently of the number of scalars 
%and fermions in the theory, 
%four-point amplitudes with only external scalars and/or fermions do not, unless the theory is some $\cN$-extended pure sYM theory or the 
%dimensional reduction of pure YM theory in some number of dimensions. We also showed that the amplitudes of pure YM theory coupled 
%with a single fermion exhibit color/kinematics duality only in the dimensions in which the theory is supersymmetric, {\it i.e.} in $D=3, 4, 6, 10$.
%
While tree-level four-point amplitudes with at least two gluons generically obey color/kinematics duality independently of the number of scalars and fermions in the theory, four-point amplitudes with only external fermions do not, unless the theory is some $\cN$-extended pure sYM theory.

It would be very interesting explore the possibility of using the amplitudes of theories obeying color/kinematics duality to construct 
amplitudes in theories where the duality is not present; see  \cite{Johansson:2013nsa} for a related discussion.
A possible approach could be to start with a higher-dimensional theory with amplitudes  obeying the duality and carry out a 
Scherk-Schwarz dimensional reduction; the numerator factors of amplitudes of the resulting lower-dimensional theory will still exhibit 
some form of color/kinematics duality while some fields will be massive. Taking the formal infinite mass limit at the level of the integrand of 
loop amplitudes ({\it i.e.} the mass is assumed to be larger than any dimensionful regulator one might choose) one is left with a massless theory
which, while {\it a priori} needs not exhibit the duality, has amplitudes whose kinematic numerators are related to each other and to the color 
factors by the Jacobi relations of the higher-dimensional theory. 

\iffalse

\begin{figure}[htb]
\centering
\includegraphics[scale=0.55]{bad_cuts.eps}
\caption{Unitarity cuts suggesting absence of color/kinematics-satisfying representations for one-loop four-point amplitudes with at least 
two external fermions or scalars (represented by dashed lines) as well as for the two-loop four-gluon amplitudes in generic matter-coupled 
gauge theories due to nontrivial contributions from the four-scalar and/or four-fermion tree-level amplitudes.}
\label{fig:bad_cuts}
\end{figure}

At loop-level this translates into potential color/kinematics satisfying one-loop gluon amplitudes for any gauge theory coupled with scalars 
and fermions; in \cite{Nohle:2013bfa} four-gluon amplitudes with this property have been constructed. 
%
We expect however that, for other choices of external lines at one loop and for gluon 
amplitudes starting at two loops, the four-point amplitudes will not have a presentation obeying this 
duality because {\it e.g.} of nontrivial contributions from the unitarity cuts shown in fig.~\ref{fig:bad_cuts}(a) 
and \ref{fig:bad_cuts}(b), respectively (dashed lines stand for scalars of fermions). 

\fi

We have also discussed the construction of scattering amplitudes in quiver gauge theories that have an orbifold point. While in 
general they may not obey color/kinematics duality due to different couplings for different gauge group factors, they can 
be obtained from the amplitudes at the orbifold point by judiciously dressing of graphs' vertices with different couplings 
for each gauge group, following the color and R-charge flow.
It would be interesting to explore whether it is possible to endow more general quiver gauge theories (or more general gauge theories) 
with color/kinematics duality. 
A possible strategy may be to embed the theory in a larger one for which color/kinematic duality is present and decouple or project 
out the extra fields at the end of the calculation.
To this end it would be interesting to study the interplay of color/kinematics duality and spontaneous symmetry breaking.

%Certain field theories with fields in the fundamental representation can be obtained from quiver gauge theories by decoupling some 
%of the gauge group factors. By decomposing the color factors of orbifold theories written in terms of the parent gauge group structure constants 
%in color factors labeled by the daughter gauge group factors it is possible to track the contribution of one particular factor; this is in the same spirit
%as deforming quiver gauge theories off the orbifold point (or "natural line"). With such a decomposition in hand it may be possible 
%to re-dress each color factor with different coupling constants and thus ultimately decouple the desired gauge group factors.

Certain field theories with fields in the fundamental representation can be obtained from quiver gauge theories by decoupling some 
of the gauge group factors. In 
%quiver gauge 
theories with an orbifold point this can be done by first deforming them off the orbifold point 
and then taking to zero the coupling of the desired gauge group.
As discussed in the introduction, an alternative strategy is to use the defining commutation relations of the gauge group as Jacobi 
identities \cite{Henrik}. 
We will argue here that this is indeed a direct consequence of the orbifold color/kinematics duality discussed in this paper for a 
non-regular orbifold which splits off one unit of rank from the gauge group of the parent theory.
%
%An alternative strategy for obtaining a theory with fields in the fundamental representation 
While the construction is quite general, 
%
%To this end we consider a non-regular orbifold group which breaks the gauge symmetry of an $SU(N+1)$ sYM parent theory down 
%to $SU(N)\times U(1)$.  While the construction is quite general, 
%
here we illustrate it by considering a ${\bf Z}_2$ the orbifold generated by
\be 
r= \text{diag}\big( -1,-1,-1,-1 \big) \ , \qquad g = \left( \begin{array}{cc} I_N & 0 \\ 0 & -1  \end{array}  \right) \ ,
\label{eg_fundamental}
\ee
which breaks the gauge symmetry of an $SU(N+1)$ sYM parent theory down to $SU(N)\times U(1)$.
It is immediate to verify that the theory has one vector and six scalars transforming in the adjoint representation of $SU(N)$, 
one $U(1)$ vector and six additional scalars which are a singlets under $SU(N)$, together with four fundamental  and four 
anti-fundamental fermions.
To exhibit the consequence of color/kinematics duality we consider an amplitude with two external fermions, 
{\it e.g.}  ${\cal A} (1^{\psi_i} 2^{\bar \psi_j} 3^{+_a} 4^{-_b})$;  the fermions are labeled by their fundamental and 
anti-fundamental indices and the two gluons carry adjoint $SU(N)$ indices.
As illustrated in section \ref{sectrees}, the tree-level amplitude in the BCJ presentation has the same numerator factors as the corresponding amplitude in the parent theory while the color factors simply need to be dressed with the color wave-functions. In our example, a solution to 
eq.~\eqref{defv} with $r$ and $g$ in eq.~\eqref{eg_fundamental} is
\bea 
v^A_{1,i} T^A_{i'j} = \delta_{i,i' } \delta_{j,N+1} \ , \qquad 
v^A_{2,j} T^A_{ij'} = \delta_{j,j' } \delta_{i,  N+1} \ , \qquad   
v^A_{3,a} = v^A_{4,a} = \delta^A_a \ , 
\eea
where $T^A_{ij}$ are $SU(N+1)$  generators. For $A<N^2$ they are also the $SU(N)$ generators and we denote the corresponding index by $a$.
With a little algebra we can rewrite the color factors as 
%follows,
\bea 
C_s &=& v^{A'}_{1,i} v^{B'}_{2,j} v^{C'}_{3,a} v^{D'}_{4,b} \ \tilde f^{A'B'X} \tilde f^{X C' D'} = \tilde f^{abX} T^X_{ji} \ , \no \\ 
C_t &=& v^{A'}_{1,i} v^{B'}_{2,j} v^{C'}_{3,a} v^{D'}_{4,b}\ \tilde f^{A'D'X} \tilde f^{X B' C'} = - (T^aT^b)_{ji} \ , \no \\
C_u &=& v^{A'}_{1,i} v^{B'}_{2,j} v^{C'}_{3,a} v^{D'}_{4,b}\ \tilde f^{A'C'X} \tilde f^{X D' B'} = (T^bT^a)_{ji} \ . 
\eea
Remarkably, the Jacobi identity of the $SU(N+1)$ gauge group of the parent theory becomes the defining commutation relation
of the $SU(N)$ gauge group of the daughter theory:
\be
C_s+C_t+C_u=0\quad \mapsto \quad [T^a, T^b] = i f^{ab}{}_cT^c \ .
\label{comm_rel}
\ee
The numerators factors corresponding to the three terms in the relation are unchanged. 

Since the defining commutation relations of the gauge group make no reference to the initial orbifold construction, the relation (\ref{comm_rel}) 
%could in principle 
can be used as the starting point for defining color/kinematics duality in presence of field in the fundamental representation for theories
which {\em do not} have an orbifold point as well as for theories with fields in {\em arbitrary} representations of the gauge group. 
The simplest instance is QED, where the gauge group is $U(1)$ and thus the right-hand side of eq.~(\ref{comm_rel})  vanishes.  This 
would suggest that the numerator factors of two Feynman graphs contributing to {\it e.g.} $e^+e^-\rightarrow\gamma\gamma$ should be equal. 
This turns out to be the case,
\be
A_{+-+-}=-2e^2\left(\frac{n}{u} + \frac{n}{t}\right)\qquad\text{with} \qquad n=\frac{\langle23\rangle\langle 13\rangle [41]^2}{s} \ .
\ee 
Here the fermion momenta are $k_1$ and $k_2$, the photon momenta are $k_3$ and $k_4$ and we choose the reference null vectors defining 
the photon polarization vectors as $q_3=k_4$ and $q_4=k_3$.\footnote{ 
Alternatively, one may simply manipulate into this form the result obtained with a more standard choice of reference spinors ({\it e.g.} one for 
which one of the two Feynman graphs vanishes).}
Using \eqref{comm_rel} and focusing on the transformation of fields in bi-fundamental representations under a single factor of the gauge group 
it may be possible to define and use color/kinematics duality for quiver gauge theories with general matter content at least on some 
"natural line" defined by specific relations between the couplings of various gauge group factors and matter fields.

Our formulation in section~\ref{loop_orb} of color/kinematics duality for orbifolds holds in principle to all orders in perturbation theory. 
It would of course be interesting and instructive to construct higher-loop examples of amplitudes and check whether or not they exhibit the duality. 
As described there, the resulting integrand will be presented as a sum of terms each of which is the contribution 
of fields of specific representations under the $SU(4)$ part of the orbifold group running in the various loops. For pure sYM 
theories -- {\it i.e.} for orbifolds with trivial action on the gauge group of the parent theory -- contributions come only from invariant fields.

The details of the construction of BCJ representations of amplitudes of orbifold theories suggested a natural proposal for 
a double-copy construction of amplitudes in non-factorized orbifold supergravities for which the orbifold group can be embedded 
in an $SU(4)\times SU(4)$ subgroup of $SU(8)$. We illustrated it in a simple example and argued that it should hold 
as long as the amplitudes of $\cN=8$ supergravity are given by a double-copy construction. 
The field content of the theories covered by this construction is insensitive to the orbifold group action on the two gauge theory factors 
and is given by the neutral part of the tensor product of their fields.
%the tensor product of fields carrying various R-symmetry representations such that the charge of the tensor product vanishes. 
Thus, since the trivial R-symmetry representation is always part of the gauge theory spectrum, the matter content of these theories is always 
larger than the one of the corresponding factorized supergravity (with amplitudes given only in terms of $n_{i,1}$ and ${\tilde n}_{i,1}$); 
pure $\cN\le 3$ supergravities cannot be constructed this way. 
It may nevertheless be possible to enhance the double-copy formula \eqref{NonfactorizedGeneral} with additional projectors such that the 
resulting spectrum is only a subset of that of the simplest factorized supergravity with the same amount of supersymmetry.

\section*{Acknowledgments }
%%%%%%%%%%%%%%%%%%%%%%%%%%%

We would like to thank Z.~Bern, J.J.~Carrasco, L.~Dixon, H.~Johansson for discussions.
This work  is supported by the US NSF under grant number PHY-1213183 (MC) 
and by the US DoE under contract DE-SC0008745 (MC, QJ and RR). RR also thanks the 
Simons Center for Geometry and Physics and the program "Physics and Mathematics of 
Scattering Amplitudes" for hospitality during the final stages of this work.
 
 \medskip 
 
As this paper was being written up we became aware of  the upcoming paper \cite{Henrik} 
where color/kinematics duality for fields in the fundamental representation is discussed in terms 
of the commutation relations of the gauge group as well as a possible double-copy approach to 
the construction of one-loop amplitudes in pure ${\cal N}\le 3$ supergravities.

\newpage

\appendix

\renewcommand{\theequation}{\thesection.\arabic{equation}}
\setcounter{equation}{0}

\section{Short summary of notation}

Since the paper is notationally heavy, we give here a summary of our notation:

\begin{tabular}{rl} \\
$(r,g)$ & generic element of the orbifold group with $ r \in SU(4)$ and $g \in SU(N)$. \\
$\Gamma$ & orbifold group, assumed to be discrete and Abelian, and hence isomorphic to ${\bf Z}_{k}$.\\
$|\Gamma|$ & rank of the orbifold group $\Gamma$.\\
$R_i$  & representation of $\Gamma$ associated to the $i$-th external leg, \\
 & product of diagonal entries of $r$. \\
$R_l$  & representation of $\Gamma$ associated to the internal leg carrying loop momentum $l$, \\
& product of diagonal entries of $r$.\\
${\cal R}$ & set of possible $R_l$, chosen case-by-case.\\
${\cal G}_3$ & set of distinct cubic graphs.\\

\end{tabular}

\section{Solutions to eqs.~(\ref{fc}) \label{app:solvefc} }

In this appendix we solve the equations \eqref{fc} for the three $2\le n_f\le 4$ allowed numbers of fermions\footnote{$n_f=0$ is trivial while 
$n_f=1$ implies that $n_s=0$ so $\lambda=0$. The resulting Lagrangians are those of pure ${\cal N}=0$ and ${\cal N}=1$ (s)YM theories.}

For $n_f=2$ (and $n_s=2$, cf. eq.~\ref{nsnf2}), the only independent equation is
\begin{equation}
\lambda^I_{12}\bar{\lambda}^{J12}+\lambda^J_{12}\bar{\lambda}^{I12}=g^2\delta^{IJ} \ ,
\end{equation}
whose solutions are
\begin{equation}
\begin{aligned}
\lambda^1_{12}=\frac{g}{\sqrt2}e^{i\theta}
\qquad , \qquad
\lambda^2_{12}=\pm i \frac{g}{\sqrt2}e^{i\theta} \ .
\end{aligned}
\end{equation}
The phase can be eliminated by redefining the fermions, $\psi \rightarrow e^{-\theta/2}\psi$, and the two signs of $\lambda^2_{12}$ 
correspond to the two possible definitions of the complex scalar field:
\begin{equation}
\frac{i}{\sqrt2}\lambda^I_{AB}\psi^A[\phi^I,\ \psi^B]\rightarrow
\frac{ig}{\sqrt2}\epsilon_{AB}\psi^A[\frac{\phi^1\pm i\phi^2}{\sqrt2},\ \psi^B] \ .
\end{equation}
The resulting Lagrangian
\begin{equation}
\begin{aligned}
{\cal L}_{\mathcal{N}=2}=&\Tr\left[-\frac{1}{4}F_{\mu\nu}F^{\mu\nu}-D_{\mu}\bar{\phi}D^{\mu}\phi+i\bar{\psi}_A\pslash{D}\psi^A
-\frac{1}{2}g^2[\phi,\ \bar{\phi}]^2\right.\\
&\left.+\frac{ig}{\sqrt2}\epsilon_{AB}\psi^A[\phi,\ \psi^B]
+\frac{ig}{\sqrt2}\epsilon^{AB}\bar{\psi}_A[\bar{\phi},\ \bar{\psi}_B]\right]\\
\end{aligned}\label{n=2}
\end{equation}
is that of $\mathcal{N}=2$ sYM theory.

\

For $n_f=3$, $n_s=4$, the equation \eqref{fc} has 6 independent components when $I=J$:
\begin{equation}
\begin{aligned}
&\lambda^I_{12}\bar{\lambda}^{I12}+\lambda^I_{31}\bar{\lambda}^{I31}=\frac{1}{2}g^2\\
&\lambda^I_{23}\bar{\lambda}^{I23}+\lambda^I_{12}\bar{\lambda}^{I12}=\frac{1}{2}g^2\\
&\lambda^I_{31}\bar{\lambda}^{I31}+\lambda^I_{23}\bar{\lambda}^{I23}=\frac{1}{2}g^2\\
&\lambda^I_{21}\bar{\lambda}^{I31}=0\\
&\lambda^I_{12}\bar{\lambda}^{I32}=0\\
&\lambda^I_{13}\bar{\lambda}^{I23}=0\ ;
\end{aligned}\label{sss0}
\end{equation}
the repeated $I$ index is not summed over. 
The first 3 equations fix the absolute value of $\lambda^I_{12}, \lambda^I_{23}$ and $\lambda^I_{31}$ for each $I=1,\dots,4$:
\begin{equation}
\label{sss1}
|\lambda^I_{12}|=\frac{g}{2}=|\lambda^I_{23}|=\frac{g}{2}=|\lambda^I_{31}| \ .
\end{equation}
This is however inconsistent with the last three equations which require that at least two of them vanish. Thus, for $n_f=3$ the equations
\eqref{fc} have no solution.

\

For $n_f=4$, $n_s=6$ the system \eqref{fc} has the following unique solution relating $\lambda$ and its conjugate:
\begin{equation}
\label{yuk}
\lambda^I_{AB}=\frac{\rho^I}{2}\epsilon_{ABCD}\bar{\lambda}^{ICD}
\quad\text{with}\qquad
\rho^I=\frac{1}{4 g^2}\epsilon^{ABCD}\lambda^I_{AB}\lambda^I_{CD} \ ;
\end{equation}
the index $I$ in the definition of $\rho^I$ is not summed over.  The condition that $\lambda$ and ${\bar\lambda}$ are conjugates of each other 
implies that $\rho^I$ is a phase. 

The remaining freedom in the choice of Yukawa couplings drops out of the Lagrangian. To see this we define the complex scalars 
\begin{equation}
\phi^{AB}=\sum_I\frac{\sqrt{\rho^I}}{g}\bar{\lambda}^{IAB}\phi^I
\qquad
\bar{\phi}_{AB}=\sum_I\frac{\sqrt{\bar{\rho}^I}}{g}\lambda^I_{AB}\phi^I \ ;
\end{equation}
it is not difficult to see that the properties \eqref{yuk} of the Yukawa couplings imply that complex conjugation is the same as lowering of indices 
with the Levi-Civita tensor:
\begin{equation}
\frac{1}{2}\epsilon_{ABCD}\phi^{CD}=\frac{1}{2}\epsilon_{ABCD}\sum_I\frac{\sqrt{\rho^I}}{g}\bar{\lambda}^{IAB}\phi^I
=\sum_I\frac{\sqrt{\bar{\rho}^I}}{g}\lambda^I_{AB}\phi^I=\bar{\phi}_{AB} \ .
\end{equation}
In terms of the new scalar field, the scalar-fermion interaction term becomes
\begin{equation}
\frac{i}{\sqrt2}\lambda^I_{AB}\psi^A[\phi^I,\ \psi^B]\rightarrow
\frac{ig}{\sqrt2}\psi^A[\bar{\phi}_{AB},\ \psi^B]
\end{equation}
while the quadratic scalar term is
\begin{equation}
D_{\mu}\bar{\phi}_{AB} D^{\mu}\phi^{AB}
=2\sum_{I,J}\frac{\sqrt{\bar{\rho}^I}}{g}\frac{\sqrt{\rho^J}}{g}\delta^{IJ}g^2D_{\mu}\phi^I D^{\mu}\phi^J
=2\sum_{I}D_{\mu}\phi^I D^{\mu}\phi^I \ .
\end{equation}
The resulting Lagrangian,
\begin{equation}
\begin{aligned}
L_{\mathcal{N}=4}=&-\frac{1}{4}F_{\mu\nu}F^{\mu\nu}-\frac{1}{4}D_{\mu}\bar{\phi}_{AB} D^{\mu}\phi^{AB}
+\frac{g^2}{16}[\bar{\phi}_{AB},\ \bar{\phi}_{CD}][\phi^{AB},\ \phi^{CD}]\\
&+i\bar{\psi}_A\bar{\sigma}^{\mu}D_{\mu}\psi^A+\frac{ig}{\sqrt{2}}\left(\bar{\psi}_A\left[\phi^{AB},\ \bar{\psi}_B\right]
+\psi^A\left[\bar{\phi}_{AB},\ \psi^B\right]\right)\\
\end{aligned}
\end{equation}
is that of ${\cal N}=4$ sYM theory.

\end{document}